\documentclass[prx,aps,twocolumn,reprint,showpacs,floatfix,superscriptaddress,longbibliography]{revtex4-1}
\usepackage{graphicx}
\usepackage{amsmath}
\usepackage{ulem}
\newcommand{\be}{\begin{equation}}
\newcommand{\ee}{\end{equation}}
\newcommand{\bea}{\begin{eqnarray}}
\newcommand{\eea}{\end{eqnarray}}

\newcommand{\ket}[1]{|#1\rangle}
\newcommand{\bra}[1]{\langle#1|}

\usepackage{xcolor}
\newcommand\hlsupp[1]{\textcolor{black}{#1}}
\newcommand\hl[1]{\textcolor{black}{#1}}

\usepackage[modulo]{lineno}
\begin{document}
\title{Thermalization near integrability in a dipolar quantum Newton's cradle}

\author{Yijun Tang}
\affiliation{Department of Physics, Stanford University, Stanford, CA 94305, USA}
\affiliation{E.~L.~Ginzton Laboratory, Stanford University, Stanford, CA 94305, USA}
\author{Wil Kao}
\affiliation{E.~L.~Ginzton Laboratory, Stanford University, Stanford, CA 94305, USA}
\affiliation{Department of Applied Physics, Stanford University, Stanford, CA 94305, USA}
\author{Kuan-Yu Li}
\affiliation{E.~L.~Ginzton Laboratory, Stanford University, Stanford, CA 94305, USA}
\affiliation{Department of Applied Physics, Stanford University, Stanford, CA 94305, USA}
\author{Sangwon Seo}
\affiliation{Department of Physics, Stanford University, Stanford, CA 94305, USA}
\affiliation{E.~L.~Ginzton Laboratory, Stanford University, Stanford, CA 94305, USA}
\affiliation{Department of Applied Physics, Stanford University, Stanford, CA 94305, USA}
\author{Krishnanand Mallayya} 
\affiliation{Department of Physics, Pennsylvania State University, University Park, PA 16802, USA}
\author{\\Marcos Rigol}
\affiliation{Department of Physics, Pennsylvania State University, University Park, PA 16802, USA}
\author{Sarang Gopalakrishnan}
\affiliation{Department of Engineering Science and Physics, CUNY College of Staten Island, Staten Island, NY 10314, USA}
\affiliation{Initiative for the Theoretical Sciences, The Graduate Center, CUNY, New York, NY 10012, USA}
\author{Benjamin L.~Lev}
\affiliation{Department of Physics, Stanford University, Stanford, CA 94305, USA}
\affiliation{E.~L.~Ginzton Laboratory, Stanford University, Stanford, CA 94305, USA}
\affiliation{Department of Applied Physics, Stanford University, Stanford, CA 94305, USA}

\date{\today}

\begin{abstract}
Isolated quantum many-body systems with integrable dynamics \hl{generically do not thermalize when taken far from equilibrium}. As one perturbs such systems away from the integrable point, thermalization sets in, but the nature of the crossover from integrable to thermalizing behavior is an unresolved and actively discussed question. We explore this question by studying the dynamics of the momentum distribution function in a dipolar quantum Newton's cradle consisting of highly magnetic dysprosium atoms.  This is accomplished by creating the first one-dimensional Bose gas with strong magnetic dipole-dipole interactions. These interactions provide tunability of both the strength of the integrability-breaking perturbation and the nature of the near-integrable dynamics. We provide the first experimental evidence that thermalization close to a strongly interacting integrable point occurs in two steps: prethermalization followed by near-exponential thermalization.  Exact numerical calculations on a two-rung lattice model yield a similar two-timescale process, suggesting that this is generic in strongly interacting near-integrable models. Moreover, the measured thermalization rate is consistent with a parameter-free theoretical estimate, based on identifying the types of collisions that dominate thermalization. By providing tunability between regimes of integrable and nonintegrable dynamics, our work sheds light both on the mechanisms by which isolated quantum many-body systems thermalize, and on the temporal structure of the onset of thermalization.
\end{abstract}

\maketitle
\section{Introduction}

In classical physics, chaos and the approach to thermal equilibrium are intimately related: the irregular space-filling trajectories of a chaotic system sample all of phase space. An integrable system, on the other hand, executes simple closed orbits. Systems that are \textit{nearly} but not strictly integrable (such as the famous Fermi-Pasta-Ulam chain~\cite{FPU55}) have a rich multiple-timescale dynamics, and equilibrate extremely slowly. Classical thermalization near integrability is understood in terms of Kolmogorov-Arnold-Moser (KAM) theory~\cite{Kolmogorov54} and related concepts. Classical chaos and KAM theory are based on the notion of phase-space trajectories, whereas \textit{quantum} chaotic dynamics and thermalization are understood in terms of a different conceptual framework, involving random matrix theory and the eigenstate thermalization hypothesis~\cite{Deutsch:1991ju, Srednicki:1994dl, Rigol:2008bf, DAlessio2015, Kaufman:2016ep, Clos:2016dv}. Within this framework, there is no general theory of thermalization in near-integrable systems, though it has been widely discussed~\cite{Burkov:2007kt, Moeckel:2008ca, Rigol:2009ew, rigol_09, eckstein2009, Kollar:2011ee, Marcuzzi:2013de, rigol2014quantum, Nessi:2014cf, Babadi15, Bertini:2015gf, Bertini:2015ds, Brandino:2015ja, Nessi:2015tt, rigol2016fundamental, Bertini:2016ho, Langen:2016bu, Biebl:2017do}. Moreover, numerical exploration of such questions is challenging because the achievable system sizes are quite small if one wishes to simulate to arbitrarily long times~\cite{Rigol:2009ew, rigol_09}.

Experimental studies are far less limited by finite-size concerns.
In a pioneering experiment~\cite{Kinoshita:2006bg}, oppositely moving bunches of ultracold bosonic atoms were confined to an array of one-dimensional (1D) tubes; atoms in this quantum Newton's cradle collided repeatedly, yet did not thermalize as atoms in a 3D trap would.  Rather than exhibiting thermalization \textit{or} revivals~\cite{FPU55}, a nonthermal momentum distribution persisted to long times. Such long-lived, nonthermal states are often termed prethermal states, and are naturally present in nearly integrable systems; they have been experimentally observed in weakly interacting, quasi-1D quantum gases~\cite{Gring:2012jd,Smith:2013hv,Langen:2015do}. The question of how such prethermal states eventually thermalize, once integrability is broken in the presence of strong interactions, remains unexplored. In particular, there is no theoretical consensus even on the basic question of whether relaxation involves two distinct timescales or three~\cite{eckstein2009,Kollar:2011ee,Nessi:2014cf,Babadi15, Bertini:2015gf}.

\begin{figure*}[t!]
\includegraphics[width=\textwidth]{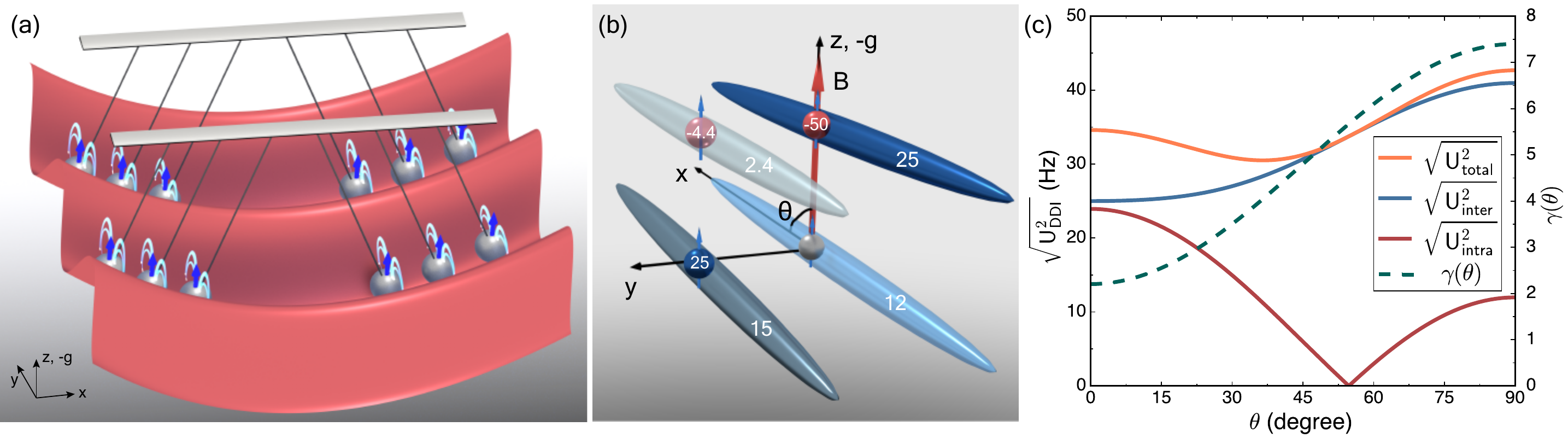}
\caption{ Dipolar Newton's cradle setup.  (a) Cartoon of two of the 1D tubes of atoms in a dipolar quantum Newton's cradle.  Application of an optical phase grating (not shown) kicks atoms along the tubes and the weak harmonic confinement induces periodic collisions.  The highly magnetic dysprosium atoms (silver spheres) are trapped in a 2D optical lattice (red) defining the tubes. (b) DDI strength between the central atom and atoms in neighboring tubes \hl{in one quadrant of the square lattice} when $\theta=90^\circ$, \hl{where $\theta$ is the angle between the B-field in the $xz$-plane and the axis of the 1D tubes along $\hat{x}$.}  \hl{The pattern of strengths in the other three quadrants are the same by symmetry.}  \hl{The DDI strength is tunable via changing $\theta$.} Numbers labeled on the atoms and the tubes are pair-wise DDI strength and integrated DDI along the tubes, respectively, in Hz.  Blue indicates large positive, while red large negative strength. (c) Dependence of integrability-breaking dipolar interaction strength and $\gamma$  on $\theta$. Solid curves: Integrability-breaking contributions of the DDI energy $U_\text{DDI}$ (added in quadrature and defined in Appendices~\ref{1DDDI}--\ref{totalDDI}) versus $\theta$. Shown are the total, total intertube, and total integrability breaking intratube DDI energies.  While the intratube DDI is maximally repulsive (attractive) for $\theta = 90^\circ$ ($0^\circ$), it vanishes among intratube atoms for $\theta = 55^\circ$.  Integrability-breaking DDI contributions come not just from the intratube 1D DDI along $\hat{x}$, but also from the 3D DDI between atoms in all neighboring tubes along $\hat{y}$ and $\hat{z}$.  This dilutes the tunablility of the DDI, reducing the contrast to a factor of $\sim$1.5 between $\theta = 0^\circ$ and $90^\circ$. Dashed curve: Lieb-Liniger parameter $\gamma(\theta)$; see Appendix~\ref{tonks} for calculation.}
\label{Fig1}
\end{figure*}

Motivated by these findings, we explore the onset of thermalization in a nearly integrable, strongly interacting system---an array of dipolar quantum Newton's cradles consisting of dysprosium atoms---subject to an integrability-breaking perturbation of \textit{tunable} strength, namely the magnetic dipole-dipole interaction (DDI); see Fig.~\ref{Fig1}(a)~\footnote{\hl{As discussed in Sec.~\ref{tunablity}, the atomic trap also breaks integrability, but only weakly compared to the DDI.}}. The tunability of our system enables us to systematically map out how the dynamics of observables changes as the system moves away from integrability; this has never before been done experimentally. We focus on an observable, the momentum distribution of the interacting dysprosium atoms, that exhibits nontrivial dynamics even in the integrable limit (because of the presence of contact interactions and confining potentials). We find that the dynamics of the momentum distribution exhibits two temporal regimes: rapid dephasing followed by a nearly exponential approach to the thermal distribution. This is similar to numerical results obtained in weakly interacting systems near a noninteracting limit~\cite{Bertini:2015gf, Bertini:2016ho} even though our integrable limit is strongly interacting. We corroborate the generality of these findings using exact diagonalization calculations of a two-rung hard-core boson model with inter-rung nearest neighbor interactions.

Furthermore, the thermalization rate extracted experimentally can be quantitatively captured by a simple physical picture: the thermalization mechanism involves an effective three-body collision, consisting of an intratube $s$-wave scattering event (the strength of which controls the parameters of the integrable model) together with an intertube dipolar scattering event (which serves as the dominant integrability-breaking perturbation). Both couplings are sensitive to the DDI. Based on our experimental observations, we argue that the thermalization rate depends not only on the strength of the integrability-breaking perturbation, but on the parameters of the integrable model itself.

\section{Dipolar quantum Newton's cradle}

The dipolar quantum Newton's cradle consists of ultracold bosonic dysprosium atoms, which have a magnetic DDI ${\sim} 100 {\times}$ stronger than, e.g., Rb's. \hl{The Bose-Einstein-condensed (BEC) atoms are tightly confined in 1D potentials created by a 2D optical lattice. The atoms are kicked into motion using an optical phase grating, and two packets of atoms in opposite momentum states collide twice each period of motion along the weakly confined direction of the 1D tubes.  The integrability-breaking interaction strength mediated by the DDI is tuned by changing the angle \hl{ $\theta$ that the dipoles of the atoms (set by the magnetic field orientation)} make with respect to the 1D tube axis; see Fig.~\ref{Fig1}(b).  We now describe the experimental details.}

\subsection{BEC production}
\label{BEC}
\hlsupp{We follow the procedure in Ref.~\cite{Tang:2015it} to produce a BEC of 1.5(2)$\times10^{4}$ $^{162}$Dy atoms in the Zeeman sublevel $m_J = -8$ ($J =8$), the absolute ground state, by evaporatively cooling in a 3D trap formed by a pair of 1064-nm optical dipole trap (ODT) laser beams crossing in $\hat{y}$ and $\hat{z}$. The $\hat{y}$-ODT beam is elliptical, with a horizontal waist of 65~$\mu$m and a vertical waist of 35~$\mu$m. The $\hat{z}$-ODT beam has a circular waist of $75$~$\mu$m. The final trap frequency before turning on the 2D optical lattice is $[\omega_x,\omega_y,\omega_z]=2\pi\times[57(1),16(2),92(2)]$~Hz.}

\subsection{2D optical lattice}

 \hlsupp{We adiabatically load the BEC into the 2D lattice by simultaneously turning on the two lattice beams using a 150-ms exponential ramp. The atoms are confined to $\sim$700 parallel one-dimensional (1D) tubes, with $\sim$50 atoms in each tube.  The 2D lattices are formed by retroreflecting a pair of beams in the $\hat{y}$ and $\hat{z}$ directions. Both beams are red-detuned from the Dy narrow-line $\lambda = 741$-nm transition~\cite{Lu:2011gc} by 13.7~GHz. The waist radii of the $\hat{z}$-lattice beam and the $\hat{y}$-lattice beam at the BEC position are 195~$\mu$m and 150~$\mu$m, respectively. Both beams are linearly polarized, and the polarization direction is chosen to be perpendicular to the applied magnetic field (confined to the $xz$-plane) such that the total AC Stark shift is maximal, including the tensor shift~\cite{Kao:17}. The $\hat{z}$-lattice beam is polarized along $\hat{y}$, such that the total light shift is constant for any $\theta$.  The polarization of the $\hat{y}$-lattice beam lies in the $xz$-plane and is set by a half waveplate to be perpendicular to the field direction for each $\theta$ setting. The lattice depth is calibrated using the Kapitza-Dirac diffraction method~\cite{Gould:1986}. We experimentally verified that the depth of the $\hat{z}$ lattice is independent of $\theta$. For the $\hat{y}$ lattice, we experimentally find the optimum waveplate angle and calibrate the lattice depth for each $\theta$ setting.} 
 
\hlsupp{We used a lattice depth of $V_0=18.0(3) E_\text{R}$, leading to a transverse trap frequency $\omega_\perp=k_\text{R}\sqrt{2V_0/m}=2\pi\times 19.0(2)$~kHz~\cite{Morsch:2006}, where $k_\text{R}=2\pi/\lambda$ is the recoil momentum and $E_\text{R}=(\hbar k_\text{R})^2/2m$. To achieve $V_0=18E_\text{R}$, the power of the $\hat{z}$-lattice beam is set to 250~mW and that of the $\hat{y}$-lattice beam is tuned between 130-170~mW as $\theta$ is changed.  This power tuning is required to compensate for both the $\theta$-dependent change in the tensor part of the atomic light shift and the loss of power through polarization-dependent optics as the laser's polarization is rotated to follow $\theta$.  The Gaussian intensity profile of the lattice beams, though broader than the ODTs, increases $\omega_x$ to $2\pi\times 60(1)$~Hz at $\omega_\perp=2\pi \times 19.0(2)$~kHz. The atoms oscillate within each tube with a frequency $1/T = 60(1)$~Hz.}

\subsection{Kicking the cradle in motion}\label{kicking}
  
\begin{figure*}[t!]
\includegraphics[width=\textwidth]{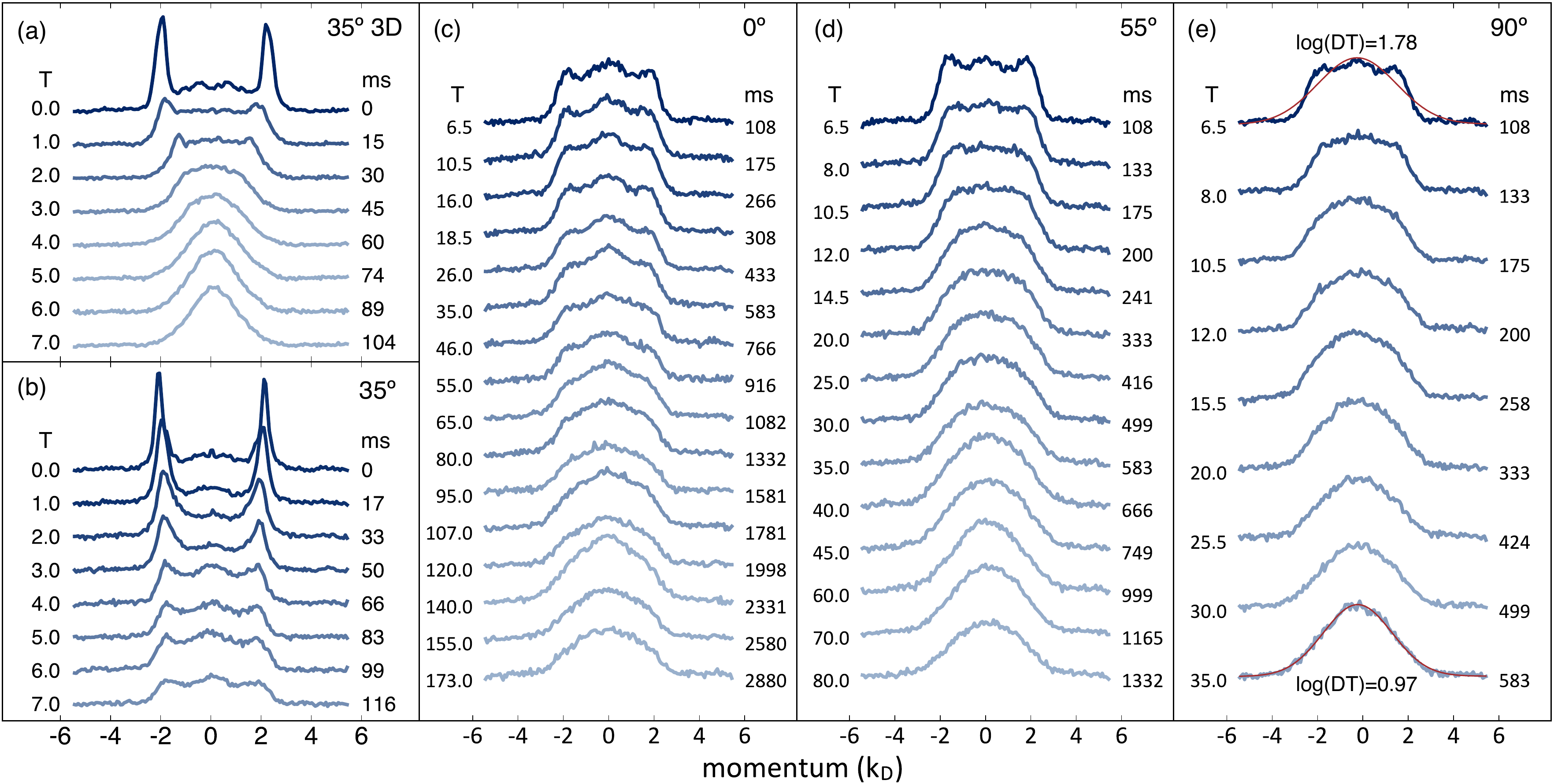}
\caption{Evolution of post-kick momentum distributions at multiples of $T$. (a) 3D gas at $\theta = 35^\circ$.  (b) Regime I, fast dephasing of 1D gas at $\theta = 35^\circ$.  While the momentum distribution of the 3D gas thermalizes after $\sim$5$T$, the 1D gases exhibit nonthermal (i.e., non-Gaussian) distributions far longer.   1D gases in regime II for $\theta$'s of (c) $0^\circ$; (d) $55^\circ$; and (e) $90^\circ$. We diffract and  evolve with $\theta= 35^\circ$ until rotation at $5T$ to the target $\theta$. As can be seen in panels (c)--(e), this procedure produces nearly identical momentum distributions after field rotation regardless of $\theta$.  Color scale is proportional to the distance to thermalization. The best-fit Gaussian curve and the corresponding $\log{(\text{DT})}$ value are shown for the $90^\circ$ data at the earliest and latest times. }
\label{Fig2}
\end{figure*}

\hlsupp{After loading into the 2D lattice, we split the gas into two equal but opposite $|{\pm} 2\hbar k_\text{D} \rangle$ momentum states by applying a precisely timed double-pulse 1D optical phase grating along the tube direction~\cite{Kinoshita:2006bg,Wu:2005,Burdick:2016njp}; $k_\text{D}=\sqrt{2}\pi/\lambda$. The phase grating beams are also red-tuned 13.7~GHz from the 741-nm transition. The two beams are linearly polarized along $\hat{z}$ and are oriented along $(\hat{x}+\hat{y})/\sqrt{2}$ and $(-\hat{x}+\hat{y})/\sqrt{2}$ directions.}
\hlsupp{Large momentum collisions can occur every $T/2=8.3(1)$~ms; the maximum collision energy between a pair of atoms is up to $E_\text{c}=2(2\hbar k_\text{D})^2/(2m)=h \times 9.0$~kHz.  This energy is three-times-lower than that required for transverse motional excitations due to the large transverse trap frequency $\omega_\perp/2\pi = 19$~kHz~\footnote{At $V_0 = 18E_\text{R}$, the excitation energy to the second excited band is 27.4~kHz~\cite{Moore:2004ka}.}.  Atomic motion is therefore restricted to 1D~\cite{Moore:2004ka, Yurovsky:2008ft}.}

\hlsupp{We experimentally observe that kicking the gas at different $\theta$ leads to different populations of undiffracted atoms.  These atoms are manifest in the momentum distribution as a small central peak in the dephased momentum distribution. This central peak,  though small, has a shape and height that varies with $\theta$ and therefore biases the distance-to-thermalization  (DT) metric of the dephased distribution. (DT is defined in Sec.~\ref{DTmetric} below.)  Among the reasons for this effect may be the dependence on the shape of the initial momentum distribution on $\theta$ due to a  dependence of the diffraction efficiency on DDI strength. To mitigate this systematic}, after kicking the gas, we allow the distribution to evolve with $\theta$ fixed to $35^\circ$ for 5 periods of oscillation before we rotate the field to the desired $\theta$ setting. This rotation takes 20~ms using a linear ramp. The ramp time is much shorter than the thermalization timescale of interest.  Appendix~\ref{rotation} shows data demonstrating that this procedure results in a dephased momentum distribution that exhibits no systematic variation in DT versus the target $\theta$ setting.  Moreover, data are shown that demonstrate that the time chosen for the rotation also does not affect the subsequent thermalization rate.

\subsection{Thermalization tunability}\label{tunablity}

To control thermalization, we break integrability through collisions mediated by the angle-tuned DDI. The effect of the DDI can be understood perturbatively as follows. In 1D, two-particle collisions only swap momenta between particles, leaving the overall momentum distribution invariant. In integrable systems, three- and more-particle collisions also have the same property: they are ``non-diffractive.'' The non-zero-range DDI breaks integrability by inducing diffractive three-particle collisions, which simultaneously change three momenta; the three particles involved need not reside in the same tube.  For example, two particles in the same tube can collide via short-range interactions while interacting with a third particle in a nearby tube via the long-range DDI.  This should lead to thermalization of the momentum distribution~\cite{Yurovsky:2008ft}. 

The DDI's \textit{anisotropic} nature, proportional to $1-3\cos^2(\theta)$, provides control of the DDI strength through tuning of $\theta$; see Figs.~\ref{Fig1}(b) and~\ref{Fig1}(c) and Appendices~\ref{1DDDI}--\ref{totalDDI}. Several experimental imperfections can also break integrability, though none in the strongly $\theta$-dependent fashion we observe.  Chief among these are heating and atom loss from spontaneous emission due to absorption of the optical trap confinement light~\cite{Johnson:2017ur}. Neither of these effects dominates thermalization at the employed trap depth; see Appendices~\ref{atomnumber} and~\ref{heating}.  Tunneling between the tubes also breaks integrability; however, we estimate its contribution to the observed thermalization is negligible; see Appendix~\ref{tunneling}. Lastly, virtual excitation of transverse motion can mediate diffractive three-body interactions and the longitudinal confinement can break integrability. Both contributions are expected to be small for our system~\cite{Mazets:2008cp, Mazets:2010hk, Tan:2010ha, Mazets:2011ko}.

\hl{We note that dipolar effects were far weaker in the Rb-based experiment of Ref.~\cite{Kinoshita:2006bg}. Dy has a dipole moment $\mu$ that is 10 times larger than Rb's. Since the thermalization rate is proportional to the dipolar interaction squared (as we demonstrate in Sec.~\ref{regime2}), and therefore to  $\mu^4$, the contribution to the thermalization rate due to  dipolar interactions was ${\sim}10^4$-slower in the Rb experiment.}

\subsection{Oscillation evolution and observation of momentum distribution}

\hlsupp{After we allow the state to dephase following the initial kick, we rotate the field to the target angle $\theta$ and hold constant the power of the lattice beams and the optical dipole trap beams for a duration of varying integer multiples of oscillation half-periods, $T/2$. To measure the evolved momentum distribution along $\hat{x}$, we first deload the lattice using a 500-$\mu$s exponential ramp, and then suddenly turn off (in $<$10~$\mu$s) the ODT beams.  The lattice deloading time is slow compared to the band-excitation timescale ($\sim$50~$\mu$s), but fast compared to the thermalization timescale in the 3D trap [$\sim$100~ms, see 3D thermalization data in Fig.~\ref{Fig2}(a)]. Therefore, this deloading procedure constitutes a band-mapping operation~\cite{Greiner:2001} that adiabatically transfers the quasimomentum distributions in the lattice confinement directions ($\hat{y}$ and $\hat{z}$) into real momentum distributions, but does not affect the momentum distribution along the tube direction $\hat{x}$, the direction of interest. }

\hlsupp{We image the gas along $\hat{y}$ after $14$~ms of time-of-flight using absorption imaging at the 421-nm transition. The images are the sum of the contributions from all tubes.
We integrate the 2D distribution along $\hat{z}$ to obtained a 1D distribution $p(x)$ because the momentum distribution of interest is along $\hat{x}$ and the band-mapping procedure produces an approximately flat distribution along $\hat{z}$ within the first Brillouin zone.}

\hlsupp{We observe no atomic population outside the lowest, ground-state band in $\hat{z}$, verifying that the 2D lattice confinement realizes an effective 1D environment for the atoms. We cannot directly observe the expanded atomic distribution along $\hat{y}$, the imaging direction, but we expect atoms also remain in the ground band due to the identical depth and deloading procedure used for both lattices. We also note that a time-of-flight expansion without transverse 1D confinement also eliminates complications arising from interaction effects during expansion. For measuring thermalization in a 3D trap as in Fig.~\ref{Fig2}(a), we diffract the BEC without loading into the lattice and allow the gas to evolve in the crossed ODT. The oscillation period in the $\hat{x}$-direction is 14.8(1)~ms in this trap. The 3D gas thermalizes within seven oscillation periods.}

\subsection{Distance-to-thermalization metric}\label{DTmetric}  
  
Figure~\ref{Fig2} shows the momentum distribution evolution of a kicked gas in a 3D dipole trap as well as the evolution for 1D gases at different $\theta$'s.  \hlsupp{We quantify the distance-to-thermalization (DT) of a measured momentum distribution $p(x)$ by fitting $p(x)$ to a Gaussian distribution $f(x)=a e^{-x^2/(2\sigma^2)}+mx+b$, where the last two terms accounts for background gradient and offset of the image, respectively. We then compute the quadrature sum of the fit residuals, $DT(t)=\sqrt{ \sum_i{ [ p(x_i) - \hat{p}(x_i) ] }^2 }$, where $\hat{p}(x_i)$ is the fitted distribution, $i$ is the pixel index, and $t$ is the holding time.} \hl{See  Appendix~\ref{heating} for a discussion of the spontaneous emission heating analysis and Appendix~\ref{dataanalysis} for comments regarding other DT metrics.}

\hlsupp{The detection noise causes $DT(t)$ to decrease to a finite positive value rather than zero when $p(x)$ becomes thermal: At long holding times $DT(t)$ reaches a constant, as evident in Figs.~\ref{Fig3} and~\ref{fig:waiting_time_comparison}. We use the mean and standard deviation of all the $DT(t)$ values in the constant region across all measurements as the mean and uncertainty of the noise floor, respectively.} The natural log of the DT is plotted in Fig.~\ref{Fig3} for these $\theta$'s.

\subsection{Interaction regime of Lieb-Liniger model}\label{liebliniger}

Pre-kick, the gas is just below the strongly correlated, Tonks-Girardeau (TG) regime of the Lieb-Liniger model \hl{wherein the bosons fermionize~\cite{Kinoshita:2005iz}}.  This regime arises when $\gamma(\theta)= mg^\text{total}_\text{1D}(\theta) / (n_\text{1D}\hbar^2)$, which contains the ratio of  the short-range (contact) interaction \hl{strength} [$\propto g^\text{total}_\text{1D}(\theta)$] to kinetic energy grows larger than unity~\cite{Olshanii:1998jr,Paredes:2004fp,Kinoshita:2004jp,Kinoshita:2005iz,Haller:2009jrb,Lang:2017cl}.  The initial $\gamma(\theta)$ varies between 0.6--1.9, where $n_\text{1D}$ is the 1D atomic density.  The unusual angle dependence of $\gamma$ arises due to the short-range, delta-function aspect of the intratube DDI; see Appendices~\ref{1DDDI} and~\ref{tonks}.

The post-kick dephasing of oscillations (which constitute regime I of evolution discussed below) reduces the initial density, allowing the gas to achieve a larger $\gamma(\theta) = 2.2$--7.4,  \hl{placing the system in the crossover to the TG}; see Fig.~\ref{Fig1}(c) for a plot of $\gamma(\theta)$ and Appendix~\ref{tonks} for more details. However, the post-kick kinetic energy scale is also much larger, \hl{and whether fermionization transiently persists during the far-from-equilibrium, post-kick evolution is a priori unclear~\cite{Kheruntsyan:2003dk}.  Once thermalized, the gas is classical in nature.} 

One can estimate post-kick interaction effects as follows: the characteristic length-scale of the nonequilibrium state is given by the wavelength of the standing-wave phase-grating pulse: $\lambda' = \lambda/\sqrt{2} \approx 520$~nm. A dimensionless ratio of this scale to $a^\text{total}_\text{1D}(\theta) = 2\hbar^2/[mg^\text{total}_\text{1D}(\theta)]$, defined as $\gamma' \equiv 2\lambda'/a^\text{total}_\text{1D}(\theta)$, generalizes the zero-temperature quantity $\gamma$ to this far-from-equilibrium situation. 
The logic is the same as when defining $\gamma$, or generally when considering whether a problem involves weak or strong correlations: one considers the ratio of the interaction strength~\footnote{\hl{Specifically, this is the bare interaction scale, or the interaction energy that would be obtained for an uncorrelated state, and \textit{not} the expectation value of the interaction in the true correlated ground state. For example, the interaction strength in the extreme Tonks-Girardeau limit approaches infinity, but the particles compensate by perfectly avoiding one another, so the interaction term has zero expectation value.}} to kinetic energy. However, since the system is far from equilibrium, the kinetic energy is no longer set by the density, but is in general much larger.
We find that $\gamma'(\theta)$ ranges from $0.9$ to $3.1$. 

\section{Thermalization observations}

\begin{figure}[t!]
\includegraphics[width=1\columnwidth]{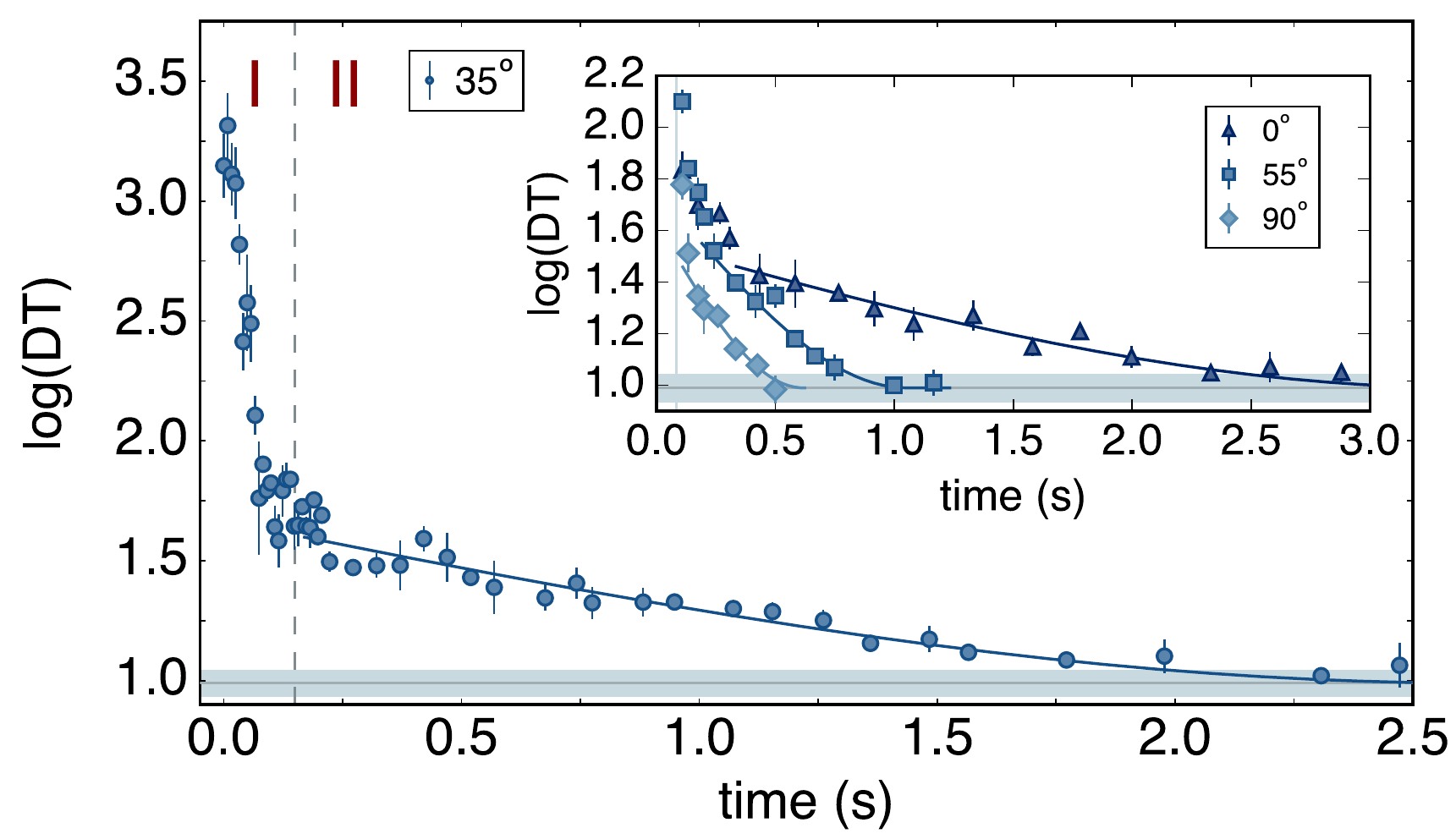}
\caption{The full  $\theta = 35^\circ$ evolution showing the boundary between regime I and regime II. Solid blue line is fit to the data between  the beginning of regime II [$\log{(\text{DT})}\approx1.5$] and the noise floor [$\log{(\text{DT})}=1$]. Vertical bars indicate standard error. Light blue horizontal band is the standard uncertainty of the noise floor. Inset contains regime II decay results for the same angles as in Fig.~\ref{Fig2}(c)--\ref{Fig2}(e).} 
\label{Fig3}
\end{figure}

We now describe the two regimes of thermalization evolution in the experimental results. The evolution of the kicked, bimodal distribution to a dephased, flattop distribution at a time $7T$ is shown in Fig.~\ref{Fig2}(b) for the example of $\theta = 35^\circ$.  Figure~\ref{Fig3} shows the full evolution for this $\theta$, where the vertical dashed line at $10T$ demarcates the boundary between regime I and II. \hl{ See Appendix~\ref{rotation} for more details.}

\subsection{Regime I evolution}\label{regime1}

\hl{The first regime, characterized by a fast decay in $\log(\text{DT})$, is governed by dephasing effects, which brings the system to a prethermal state. Dephasing is dominated by dynamics arising from the inhomogeneous trapping potential in the presence of interactions. There are two distinct dephasing processes due to the trap: (1) dephasing of oscillations between different harmonic tubes, owing to their different natural frequencies and subsequent ensemble averaging over tubes with different $T$ during the imaging process; and (2) dephasing of the oscillations of the gas in a \textit{single} tube, owing to its anharmonicity. Both processes were discussed in Ref.~\cite{Kinoshita:2006bg}. These processes correspond to different physics: process (1) yields an approximately stationary state as an artifact of averaging over tubes, while process (2) causes dephasing in each individual tube.} 

\hl{We have quantified these trap-induced technical dephasing processes by a collisionless classical simulation of the particle dynamics in each anharmonic tube, averaged over the inhomogeneous tubes; see Appendix~\ref{collisionless}. This allows us to use knowledge of trap parameters to predict the dephasing timescale of processes (1) and (2).  The simulation shows that the momentum distribution completely dephases in approximately 150~ms. It also shows that the contribution from process (2) is as important as process (1).}

\hl{We note that the simulated dephasing time is slightly longer than that observed in the experiment; see Fig.~\ref{Fig2}(b). The discrepancy is likely due to the lack of interactions in the simulation, though could also be the result of an imperfect modeling of the trap arising from uncharacterized distortions to the beam shapes and overlap of the beam focus.  Interactions are expected to rapidly broaden the initial momentum peaks~\cite{vandenBerg:2016cs} and, hence, to speed up dephasing.  Indeed, our exact diagonalization simulations in Fig.~\ref{Fig5} show that a rapid decay due to interacting integrable dynamics ensues after the quench even in the absence of technical dephasing. (Section~\ref{simulations} and Appendix~\ref{XXZcalc} describe these simulations in more detail.)  
However, we remark that in a strictly harmonic trap, integrable interactions alone are not expected to yield a stationary distribution as we observe in our anharmonic system: processes (1) and (2) together with interactions are important in the experiment.}

\hl{To gain further understanding of the interplay between technical dephasing and interaction effects in our system, we performed an experimental study involving a \textit{single-sided} kick measurement.
This measurement reveals the effect of technical dephasing in the absence of high-energy collisions, i.e., head-on collisions. See Appendix~\ref{dephasing} for  experimental details and comments. We observe that the dephasing time, where the initial fast decay of the DT transitions to a much slower decay, is $\sim$70~ms and is similar to the dephasing time observed in our double-sided data in Fig.~\ref{Fig2}(b).  We conclude that dephasing due to anharmonicity and inhomogeneity in the presence of interactions, but in the absence of large-momentum collisions, explains regime I evolution.  }

\hl{We note that integrable dynamics immediately after a quench are often referred to as prethermalization.  During prethermalization, observables not directly related to the conserved quantities dephase; see, e.g., Refs.~\cite{Berges:2004ef, Gring:2012jd, Langen:2016bu}.  In this experiment, the observable is the momentum distribution of the atoms, while what is conserved at integrability is the distribution of the so-called rapidities. At zero density in interacting systems, or in noninteracting systems, the rapidities are the same as the momenta of the particles (in systems that are translationally invariant). However, at nonzero densities in interacting systems the rapidities are not easily related to the momenta of the atoms~\cite{giamarchi2003quantum}. As a result, even though the distribution of rapidities does not change, physical observables such as the momentum distribution function of the atoms can change and do change in this experiment.}

\hl{ Regime I is the prethermalization regime in our experiment. The dephased state at the end of regime I can be described using a generalized Gibbs ensemble~\cite{Rigol:2007bm} arising from the combination of all three mechanisms.  Physically, mechanisms (1) and (2), and integrable interactions, give rise to a dephased state through independent prethermalization processes, and each dephased state may be described by a generalized Gibbs ensemble. We therefore refer to the final dephased state at the end of regime I as a ``prethermal state'' regardless of its prior history. We now turn to the thermalization of this prethermal state.}

\subsection{Regime II evolution and thermalization rate}\label{regime2}

\begin{figure}[t!]
\includegraphics[width=\columnwidth]{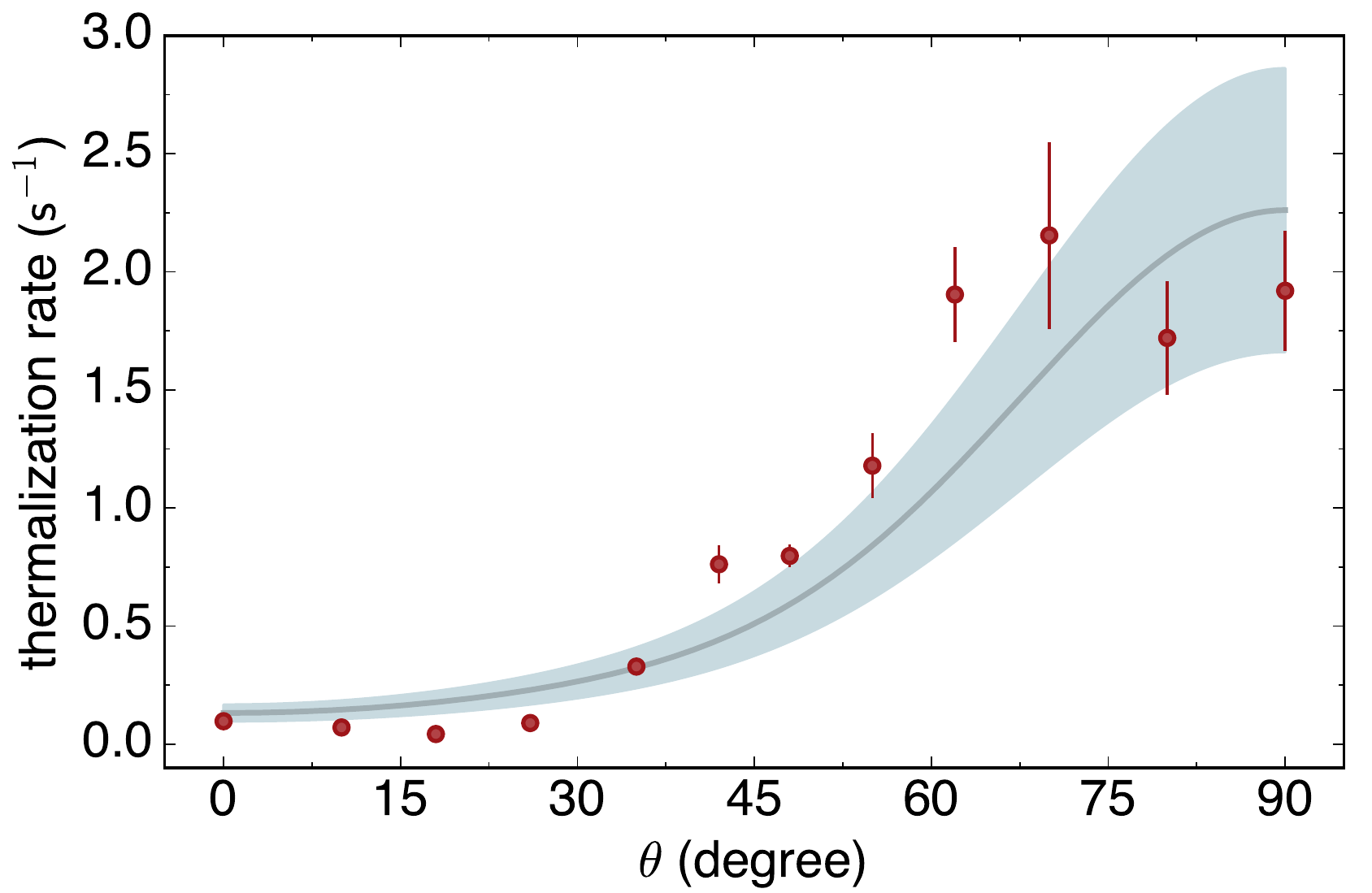}
\caption{ Thermalization rate data in red versus $\theta$. See Eq.~\ref{eq1} for definition of thermalization rate $1/\tau_\text{th}$.  Gray curve is the scaling estimate  $\gamma'^2(\theta)U^2_\text{total}(\theta)/(E_\text{c})$ with no free parameters or offset. Vertical bars and light blue band indicate standard error;  atom number noise and $a_\text{3D}$ uncertainty dominate the latter.  Evidently, the thermalization rate in the dipolar quantum Newton's cradle is well-described by terms dependent on both the long-range  and short-range parts of the  DDI, the former through total (inter- plus intratube) $U_\text{total}^2(\theta)$ and the latter through the intratube DDI dependence of $a^\text{intra}_\text{1D}(\theta)$ in $\gamma'^2(\theta)$. }
\label{Fig4}
\end{figure}

\hlsupp{To determine the thermalization rate for the regime II slow-decay evolution data shown in Fig.~\ref{Fig3}, we fit the regime II decay to
\begin{equation}\label{eq1}
\log{\text{DT}}=\begin{cases}
\sqrt{\left(\log{\text{DT}_0}\right)^2 + [(t-t_\text{th})/\tau_\text{th}]^2}, & t<t_\text{th}\\ 
\log{\text{DT}_0}, & t \ge t_\text{th}
\end{cases},
\end{equation}
which is asymptotically a single exponential decay characterized by a rate $1/\tau_\text{th}$ at short times and becomes a constant noise value $\text{DT}_0$ at long time. Here $\tau_\text{th}$ and $t_\text{th}$ are free parameters, and $\text{DT}_0$ is determined from the data. }

The fitted rates  versus $\theta$ are plotted in Fig.~\ref{Fig4}. \hl{The rates are corrected for spontaneous-emission heating; see Appendix~\ref{heating}.}  Comparing to the total DDI and $\gamma(\theta)$ plotted in Fig.~\ref{Fig1}(c), we see that the slowest (fastest) thermalization rate occurs at small (large) $\theta$ where both the DDI and $\gamma$ are smallest (largest), with a monotonic increase from low to high $\theta$. While there is no ab initio theory we can yet invoke to explain either this trend or magnitude, we can provide a simple estimate. We expect the thermalization rate to scale as the square of both the contact and dipolar interactions, as the largest integrability-breaking perturbation involves both an $s$-wave collision and a two-body dipolar collision. The matrix element giving rise to thermalization is linear in both the $s$-wave collision rate and the DDI, and thus by Fermi's Golden Rule, the rate is quadratic in both quantities. An appropriate measure of contact interactions in the nonequilibrium state is $\gamma'$, as argued in Sec.~\ref{liebliniger} above. This suggests that the rate should scale as $\gamma'^2(\theta) U^2_\text{total}(\theta)/E_\text{c}$, where $E_\text{c} = 2E_\text{k}$ is the collision energy of two intratube atoms and $E_\text{k} = (2\hbar k_\text{D})^2/2m$. This simple estimate, plotted in Fig.~\ref{Fig4}, is in good quantitative agreement with the data.

\section{Exact diagonalization calculations}\label{simulations}

In what follows, we relate the observation of a two-timescale evolution to the dynamics obtained in exact diagonalization calculations of a two-rung lattice model of hard-core bosons.

\subsection{Setup}

\begin{figure}[t!]
	\centering
	\includegraphics[width=0.45\textwidth]{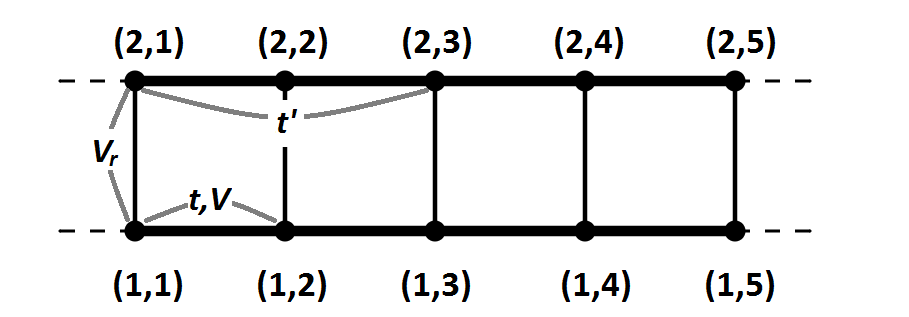}
	\caption{\hlsupp{Two-rung lattice made of two identical chains with nearest neighbor hopping ($t$), interaction ($V$), and next-nearest neighbor hopping ($t'$). The two chains interact along the rungs ($V_r$).}}\label{2runglattice}
\end{figure}

\hlsupp{The lattice consists of two identical 1D chains, with each chain described by a $t$-$t'$-$V$ Hamiltonian with nearest neighbor hopping $t$, next-nearest neighbor hopping $t'$, and nearest neighbor interaction $V$. The two chains interact along the rungs, with a strength set by $V_r$, to mimic the intertube DDI in the experiment (see Fig.~\ref{2runglattice}). The Hamiltonian can be written as
\begin{eqnarray}\label{Hi}
\hat{H}&=&\sum_{\ell=1}^{2}\sum_{i=1}^{L/2}  -t\left( \hat{b}_{\ell,i}^{\dagger}\hat{b}^{ }_{\ell,i+1} +\text{H.c.}\right)-t'\left( \hat{b}_{\ell,i}^{\dagger}\hat{b}^{ }_{\ell,i+2} +\text{H.c.} \right)\nonumber \\
&+& \sum_{\ell=1}^{2}\sum_{i=1}^{L/2} V\left(\hat{n}^{ }_{\ell,i}-\frac{1}{2}\right)\left(\hat{n}^{ }_{\ell,i+1}-\frac{1}{2}\right) \nonumber\\
&+& \sum_{i=1}^{L/2} V_r\left(\hat{n}^{ }_{1,i}-\frac{1}{2}\right)\left(\hat{n}^{ }_{2,i}-\frac{1}{2}\right),
\end{eqnarray}
where $\hat{b}_{\ell,i}^{\dagger}$ ($\hat{b}_{\ell,i}^{}$) is the creation (annihilation) operator at site $i$ in chain $\ell$ (=1,\,2), and $\hat{n}_{\ell,i} = \hat{b}_{\ell,i}^{\dagger}\hat{b}^{ }_{\ell,i}$ is the site occupation operator. $L$ denotes the total number of sites in the lattice, which has $L/2$ sites per chain. Periodic boundary conditions along the chains are imposed by the conditions $(\ell, L/2+1)=(\ell,1)$ and $(\ell, L/2+2)=(\ell,2)$.}

 \hlsupp{The hard-core boson creation-annihilation operators obey bosonic commutation relations  $[\hat{b}_{\ell,i},\hat{b}_{\ell',j}]=[\hat{b}_{\ell,i}^{\dagger},\hat{b}_{\ell',j}^{\dagger}]=0$, $[\hat{b}^{ }_{\ell,i},\hat{b}_{\ell',j}^{\dagger}] = \delta_{\ell,\ell'}\delta_{i,j}$, supplemented by the constraints $\hat{b}_{\ell,i}^2 = \hat{b}_{\ell,i}^{\dagger} { }^2 = 0$ to prevent multiple occupancy of the lattice sites. When $t'=V_r=0$, the Hamiltonian  reduces to that of two disconnected integrable chains (the spin-1/2 XXZ Hamiltonian in the spin language) and can be solved using the Bethe ansatz \cite{Cazalilla:2011dm}. For $V=t'=V_r=0$, the chains become the lattice analogue of the Tonks-Girardeau gas, and the Hamiltonian can be mapped onto that of noninteracting spinless fermions \cite{Cazalilla:2011dm}. Given $t\neq0$ and $V\neq0$, the Hamiltonian  is nonintegrable for $t'\neq0$ and/or $V_r\neq0$. We are mostly interested in dynamics when $V_r\ll V$ ($t'=0$), so that integrability is weakly broken.} 

\hlsupp{We take our initial states to be in thermal equilibrium, as described by the grand canonical ensemble (GE), for the initial Hamiltonian $\hat{H}_I$
\begin{equation}\label{eq:GE}
\hat\rho_I=\frac{\exp[-(\hat H_I-\mu_I \hat N)/T_I]}{\text{Tr}\{\exp[-(\hat H_I-\mu_I \hat N)/T_I]\}},
\end{equation}
where $\hat{N}=\sum_{\ell,i} \hat{n}_{\ell,i}$ is the total number of particle operator, $T_I$ is the initial temperature (we set the Boltzmann constant to 1), and $\mu_I$ is the initial chemical potential. We set $\mu_I=0$ in all our calculations, which results in the lattices being at half filling because of the particle-hole symmetry of Hamiltonian \eqref{Hi}.}

\hlsupp{The system is taken out of equilibrium by a sudden quench in which $\hat{H}_I$ is changed to $\hat{H}_F$, such that $[\hat{H}_F,\hat{H}_I]\neq 0$. The system is assumed to be isolated so that the ensuing dynamics is unitary. The density matrix at time $\tau$ after the quench is given by (we set $\hbar=1$) 
\begin{eqnarray}
\hat \rho(\tau) &=& e^{-i\hat H_F \tau}\hat\rho_I e^{i\hat H_F\tau} \nonumber\\
&=& \sum_{n,n'}e^{-i(E_{n'}-E_n)\tau}\ket{n'}\bra{n'}\hat\rho_I\ket{n}\bra{n}, \label{rho_time}
\end{eqnarray}
where $\ket{n}$ and $E_n$ are the energy eigenkets and eigenvalues of $\hat H_F$, respectively.} 

\hlsupp{Our observable of interest is, as in the experiments, the momentum distribution ($\hat{m}_k$) along the chains (see also Appendix~\ref{XXZcalc1})
\begin{equation}
\hat m_{k}=\frac{1}{L}\sum_{\ell=1}^{2}\sum_{j,j'=1}^{L/2} e^{ik(j-j')} \hat b_{\ell,j}^{\dagger}\hat b_{\ell,j'}^{ }.
\end{equation}
The time dependence of $\hat{m}_k$ is studied by computing $m_k(\tau)=\text{Tr}[\hat m_{k}\hat \rho(\tau)]$, while the expectation value of this observable after relaxation can be obtained from the infinite-time average \cite{DAlessio2015}
\begin{eqnarray}
\bar{m}_k=\lim_{\tau'\rightarrow\infty}\frac{1}{\tau'}\int_{0}^{\tau'} m_k(\tau)\, d\tau.
\end{eqnarray}
In the absence of degeneracies, which is ensured in our calculations by breaking down the Hamiltonian into its symmetry irreducible sectors, the infinite-time average agrees with the prediction of the so-called diagonal ensemble (DE) \cite{Rigol:2008bf}:
\begin{eqnarray}
m_k(\text{DE})=\sum_{n}\bra{n}\hat \rho_{\text{DE}}\ket{n}\,\bra{n}\hat m_{k}\ket{n},
\end{eqnarray}
where
\begin{eqnarray}
\hat \rho_{\text{DE}}=\lim_{\tau'\rightarrow\infty}\frac{1}{\tau'}\int_{0}^{\tau'} \hat \rho(\tau)\, d\tau.
\end{eqnarray}}

\hlsupp{A central question we address with the exact diagonalization calculations is how the momentum distribution equilibrates after the quench. For that, we compute a ``distance-to-equilibrium'' as the RMS deviation of the momentum distribution function at each time from the DE prediction:
\begin{eqnarray}\label{eq:dde}
\delta_{\text{DE}}(\tau)=\sqrt{\displaystyle\frac{\sum_k \left[m_k(\tau)-m_k(\text{DE})\right]^2}{L/2}}.
\end{eqnarray}
See Appendix~\ref{XXZcalc2} for a discussion of thermalization.}

\subsection{Numerical results}

\hlsupp{We set $t=1$ (our energy scale) before and after the quench (and set our unit of time to $\hbar/t=1$). As mentioned before, our quenches start from an initial state in thermal equilibrium. We take the temperature to be $T_I=5t'_I$ (qualitatively similar results were obtained for other temperatures) for an initial Hamiltonian $\hat H_I$ that has $t'_I=50$ and $V^I_r=0$. A large $t'$ in $\hat H_I$ is chosen to create an initial momentum distribution that peaks at $k=0$ and $k=\pi$ (see Fig.~\ref{fig:XXZmomentum}). This is done to resemble the post-kick bimodal initial state created in the experiment. After the quench, $t'$ in $\hat H_F$ is set to 0 and $V_r$ is set to various nonzero but small values, so that the evolution occurs under a (in most cases) weakly nonintegrable Hamiltonian. Exploiting translation symmetry, particle-hole symmetry, number conservation per chain in the two-rung system, as well as parity under space reflection, we perform exact diagonalization calculations in systems with up to $L=22$ sites. The value of $V$ is kept constant during the quench and is selected to be $V=1.6$.}

\begin{figure}[t!]
\includegraphics[width=1\columnwidth]{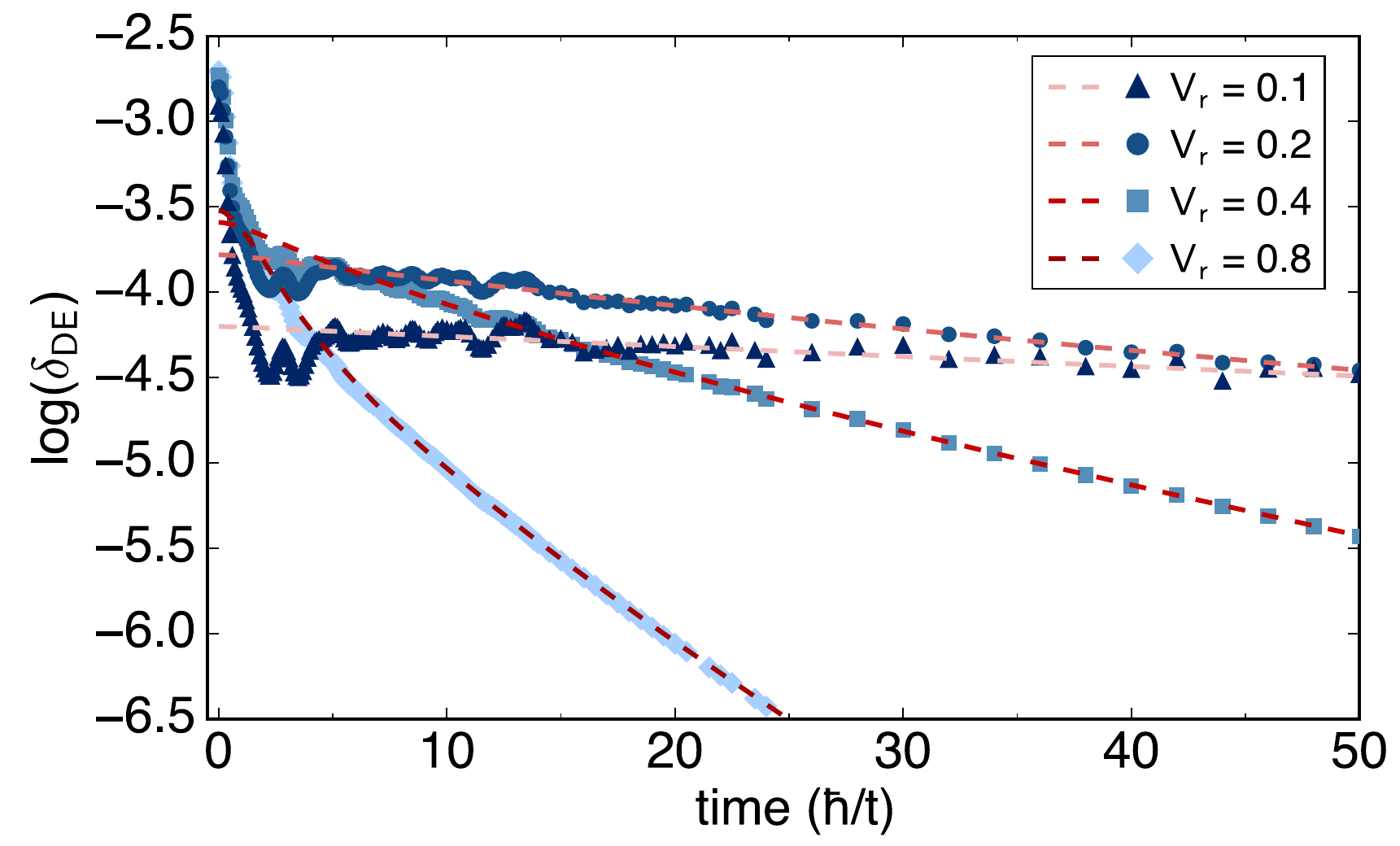}
\caption{Numerical results for the approach to equilibrium [see Eq.~\eqref{eq:dde}] in the two-rung hard-core boson model calculations with 22 lattice sites and nearest neighbor coupling $V=1.6$. The symbols show results for a quench in which the system is initialized in a state with a two-peaked momentum distribution (created through an initial Hamiltonian with strong next-nearest neighbor coupling $t' = 50$), and the integrability-breaking interaction is turned on post-quench.  Dashed lines show results for evolution under the same final Hamiltonian, but from an initial state that has already dephased under the fast integrable dynamics. Specifically, the initial state is a diagonal-ensemble state generated by a quench in which $t'=50\rightarrow t'=0$ is changed but dipolar interactions are absent. The fast dephasing at short times in the simulation depends weakly on the strength of the integrability-breaking perturbation.} 
\label{Fig5}
\end{figure}

\hl{Figure~\ref{Fig5} shows the ``distance-to-equilibrium'' $\delta_{\text{DE}}$ plotted as a function of time for four values of the strength of the integrability breaking inter-rung coupling $V_r$. Like in the experiments, one can see that the exact diagonalization results exhibit two-timescale dynamics. Prethermalization occurs for times $\lesssim \hbar/t$, a time-scale set by $V\sim t$. The near-exponential approach to the diagonal ensemble result occurs in a longer timescale, which is set by $V_r$.}

\hlsupp{The experiment strives to use the same initial state to study the approach to thermalization when the strength of the DDI (set by $\theta$) is changed. The initial state is taken to be the one after the short-time dephasing for a particular value of $\theta$ ($\theta$ is changed after that). We can emulate such a procedure in our numerical calculations by ``splitting'' our single quench in which $t'$ is set to zero, and $V_r$ is made nonzero, into a two-step quench. In the first quench, $t'$ is set to zero (this is a quench to the integrable part of $\hat H_F$) and the system is allowed to equilibrate. One can then take the equilibrated state as the initial state for a second quench in which the integrability-breaking interaction $V_r$ is turned on. Alternatively, one can take the diagonal ensemble after the first quench as the initial state for the second quench. Both procedures produce indistinguishable relaxation rates (see Appendix~\ref{appequil}).}

\hl{Using the diagonal ensemble after the first quench as the initial state for the second quench allows us to separate out the effects of the integrable and the nonintegrable parts of the Hamiltonian. The ensuing dynamics, shown as dashed lines in Fig.~\ref{Fig5}, are indistinguishable from those of the original quench after the short-time dephasing, making apparent that regime II is entirely due to the integrability-breaking interactions.}

\hl{The similarity between the results in Fig.~\ref{Fig3} and in Fig.~\ref{Fig5} is striking considering that the systems studied experimentally and theoretically are microscopically very different. By doing so, it highlights the robustness of our findings about the relaxation dynamics close to a strongly interacting integrable point \cite{Mallayya:2017vf}.}

\section{Conclusions}

In summary, we explored the far-from-equilibrium dynamics of a strongly interacting nearly-integrable system as it is systematically tuned away from integrability. We provide the first experimental demonstration that observables in such systems thermalize in a two-step process:  prethermalization followed by near-exponential thermalization. A similar behavior is observed in exact numerical calculations of a strongly interacting lattice model. We have also shown that the thermalization rate in our experiments is well-described by a DDI-dependent scaling function that is consistent with perturbative expectations: The scaling is quadratic in the effective intratube contact interactions, and also in the intra- and intertube dipolar interactions.

Our ability to control the strength of integrability-breaking perturbations opens a new venue to explore quantum thermalization in strongly interacting systems. Many questions remain, such as how thermalization depends on the ``quantumness'' of the system, which we can also control by changing the amount of energy deposited in the initial state. Our detailed characterization of the approach to the thermal regime can also play an important role in the development and benchmarking of quantum Boltzmann approaches that could be used in other areas of physics, such as heavy ion collisions. 

\begin{acknowledgments}

We thank David Weiss, Nigel Cooper, Pavan Hosur, Chao Wang, Eugene Demler, {J\"{o}rg} Schmiedmayer, and Robert Konik for enlightening discussions.  We acknowledge support from NSF (PHY-1403396, PHY-1707482, DMR-1653271, as well as PHY-1607611 via the Aspen Center for Physics), and AFOSR (FA9550-12-1-0056). WK acknowledges support from NSERC Postgraduate Scholarship-Doctoral. 

\end{acknowledgments}

% \pagebreak
% \clearpage

% \onecolumngrid
%\section*{Supplemental Materials}

%  \renewcommand\thefigure{\thesection.\arabic{figure}}  
%\setcounter{figure}{0}  

  \appendix
  
\section{Intratube dipolar interaction}\label{1DDDI} The dipole moment $\mu$ of Dy is 9.93$\mu_\text{B}$.  The effective 1D dipole-dipole interaction (DDI) has been derived in the single-mode approximation to be~\cite{Sinha:2007gx,Deuretzbacher:2010gg,Deuretzbacher:2013da}:  
   \begin{equation}
       U^\text{1D}_\text{DDI}(x) = V(\theta)\left[ {V}^\text{1D}_\text{DDI}(u) - \frac{8}{3} \delta (u) \right],
   \label{ddi_1d} 
   \end{equation}
where
 \begin{equation}
  V(\theta) = \frac{\mu_0 \mu^2}{4\pi} \frac{1-3\cos^2{\theta}}{4 l_\perp^3},
   \end{equation}
   \begin{equation}\label{vtilda}
       {V}^\text{1D}_\text{DDI}(u) = -2|u| + \sqrt{2\pi} (1+u^2) e^{u^2/2} \text{erfc}(|u|/\sqrt{2}),
   \end{equation}
and $u=x/l_\perp$,  $l_\perp = \sqrt{\hbar/m\omega_\perp}$, and  $\text{erfc}(u)$ is the complementary error function. The $\delta$-function term in Eq.~\eqref{ddi_1d} comes from the point limit of an extended dipole~\cite{Griffiths:1998gi} and has an opposite sign to ${V}^\text{1D}_\text{DDI}(u)$. For large distances $|x|\gg l_\perp$, ${V}^\text{1D}_\text{DDI}(u) \rightarrow 4/|u|^3$, just like the DDI in 3D. However, ${V}^\text{1D}_\text{DDI}(u)$ assumes a finite value at the origin,  becoming more sharply peaked for smaller $l_\perp$. This behavior resembles that of a $\delta$ function and allows one to define an effective $\delta$-function potential for ${V}^\text{1D}_\text{DDI}(u)$ at short distance~\cite{Deuretzbacher:2010gg}.

We note that the intratube DDI is suppressed as atoms approach within a few $l_\perp$ by a factor of $4/|u|^3/[2|u| - \sqrt{2\pi} (1+u^2) e^{u^2/2} \text{erfc}(|u|/\sqrt{2})]$.  To understand this reduction in 1D, consider $\theta = 90^\circ$. While most of the DDI between atoms along $\hat{x}$ is repulsive (i.e., dipoles lying abreast), there remains a small attractive contribution (i.e., dipoles lying head-to-tail) from the part of their wavefunctions that extend transversely by $l_\perp$. In general, if the DDI interaction between two dipoles is repulsive when they are separated in the longitudinal direction (side-by-side), their interaction will be attractive when separated in the transverse direction (head-to-tail), and vice versa, reducing the strength of the DDI in either case.  See Ref.~\cite{Deuretzbacher:2010gg} for details.

In the following discussions, we use the superscript to denote the interaction range (``sr" for short-range and ``lr" for long-range) and the subscript to denote the nature of the interaction (``intra" for intratube and ``inter" for intertube).

\subsection{Short-range part of the intratube dipolar interaction}\label{deltaintraDDI}

The magnitude of the short-range part of the 1D DDI is given by the sum of the term proportional to the $\delta$ function $-\frac{8}{3} \delta(u) $ in Eq.~\eqref{ddi_1d} and the $\delta$-function-like part of ${V}^\text{1D}_\text{DDI}(u)$ in Eq.~\eqref{vtilda}.  We calculate the strength of the  ${V}^\text{1D}_\text{DDI}(u)$ by integrating it over a suitably chosen spatial domain in $\hat{x}$.  Reference~\cite{Deuretzbacher:2010gg} determines this range to be $\pm\sqrt{2\pi} l_\perp$, which is sufficiently smaller than the interparticle spacing inside the tube such that the long-range $1/r^3$ tail of the DDI is not double-counted.  Taking $u \in [-\sqrt{2\pi},+\sqrt{2\pi}]$ as in Ref.~\cite{Deuretzbacher:2010gg}, we find the normalized strength $A$ of the short-range part of the interaction  ${V}^\text{1D}_\text{DDI}(u)$ to be  \begin{equation}
       A = \int_{-\sqrt{2\pi}}^{+\sqrt{2\pi}} {V}^\text{1D}_\text{DDI}(u)\,du \approx 90\%\int_{-\infty}^{+\infty}  {V}^\text{1D}_\text{DDI}(u)\,du = 3.6.
   \end{equation}
This leads to a DDI-induced $\delta$-function interaction strength \begin{equation}\label{gddi1d}
U^\text{sr}_\text{intra}(\theta) = g^\text{DDI}_\text{1D}(\theta)\delta(x) = V(\theta)(A-8/3)l_\perp \delta (x).
\end{equation}

\subsection{Long-range part of the intratube dipolar interaction}\label{longintraDDI}

The long-range (i.e., $1/r^3$-scaling) part of the intratube DDI is given by $V(\theta)B$, where
  \bea
      B &=& \int_{-\infty}^{+\infty}  {V}^\text{1D}_\text{DDI}(u)\,du - \int_{-\sqrt{2\pi}}^{+\sqrt{2\pi}}{V}^\text{1D}_\text{DDI}(u)\,du \nonumber \\ 
      & \approx & 10\%\int_{-\infty}^{+\infty} {V}^\text{1D}_\text{DDI}(u)\,du = 0.4.
  \eea
However, not all of long-range intratube DDI contributes to the integrability-breaking perturbation; only the momentum-dependent part can lead to momentum randomizing collisions.  To find the leading momentum-dependent part, we expand the Fourier transform of ${V}^\text{1D}_\text{DDI}(u)$ up to $O(k^2)$, which provides the terms associated with the DDI-induced virtual interactions leading to integrability breaking; higher-order terms would contribute less to thermalization.  The $k$-space form of the DDI is  ${V}^\text{1D}_\text{DDI}(k) \sim [1 - \sigma\exp{\sigma}\Gamma(0,\sigma)]$~\cite{Sinha:2007gx}, where $\sigma = (2 k_\text{D} l_\perp)^2/2\approx 0.2$ and $\Gamma(0,\sigma)$ is the incomplete Gamma function.   The result is $\eta = (\widetilde{\gamma}+\log{\sigma})\sigma$,
where $\widetilde{\gamma}=0.577...$ is the Euler-Mascheroni (Euler-Gamma) constant.
The integrability-breaking term from the intratube DDI is therefore $\eta V(\theta)B$.

\section{Intertube dipolar interaction}\label{interDDI}

Due to the lack of spatial correlations between atoms in nearby tubes after splitting, the intertube DDI should be calculated as that between an atom in one tube and the integral over all $x$ positions in the nearby tube.  More explicitly, for a tube located at $(y,z)$,
\begin{equation}
    U_\text{inter}^{y,z}(\vec{B}) =n_\text{1D} \int^{+\infty}_{-\infty}{V_\text{inter}(\vec{r},\vec{B})\,dx},
\end{equation}
where 
\begin{equation}
V_\text{inter}(\vec{r},\vec{B}) = \frac{\mu_0 \mu^2}{4\pi}\frac{1-3\left(\hat{r}\cdot\hat{B}\right)^2}{r^3}.
\end{equation}
Here $\vec{r} = (x,y,z)$ denotes the atomic position vector and $\hat{B}$ the direction of the magnetic field.  With the geometry of our experimental setup, we can parameterize the two vectors as
\begin{align*}
\vec{r} &= a\left(\frac{x}{a},i,j\right),\\
\vec{B} &= B\left(\cos\theta,0,\sin\theta\right),\\
V_\text{inter}^{i,j}(\theta) &= V_\text{inter}(\vec{r},\vec{B}),
\end{align*}
where $a = \lambda/2 = 371$~nm is the lattice constant and $i$, $j$ are integer indices that denote the location of each tube.

Dimer bound states are predicted to form between pairs and arrays of tubes for any negative interaction $U_\text{inter}^{y,z}(\theta)<0$~\cite{Zinner:2011cd,Volosniev:2013by}.  However, these complexes would have binding energies far lower than the post-kick atomic collision energy, and so are unlikely to survive the kicking process.  We therefore do not expect intertube spatial atomic correlations to arise from pre-kick dimer formation. 

\section{Calculation of $U_\text{total}^2(\theta)$}\label{totalDDI}

We calculate $U_\text{total}^2$, the quadrature sum of the integrability-breaking DDI contributions, using
\begin{equation}
U_\text{total}^2(\theta) = [\eta U^\text{lr}_\text{intra}(\theta)]^2 + \sum_{i,j}  \left[U^{i,j}_\text{inter}(\theta)\right]^2,
\end{equation}
where $i$, $j$ are the tube indices, and 
\begin{widetext}
\be
U^\text{lr}_\text{intra}(\theta) = V(\theta) \left(n_\text{1D} l_\perp\right)^\frac{1}{2} \sqrt{\int_{-\infty}^{-\sqrt{2\pi}} \left[V^\text{1D}_\text{DDI}(u)\right]^2\,du+\int_{+\sqrt{2\pi}}^{+\infty} \left[V^\text{1D}_\text{DDI}(u)\right]^2\,du},
\ee
\end{widetext}
\be
U^{i,j}_\text{inter}(\theta) = \sqrt{n_\text{1D}\int^{+\infty}_{-\infty}\left[V^{i,j}_\text{inter}(\theta)\right]^2\,dx_{i,j}}.
\ee
The magnitudes of the integrability-breaking intra- and intertube DDI energies, $\eta U^\text{lr}_\text{intra}(\theta)$ and  $\sqrt{\sum_{i,j}  [U^\text{i,j}_\text{inter}(\theta)]^2}$ for $i,j \le 2$, are plotted in Fig.~\ref{Fig1}(c).

\section{Lieb-Liniger parameter $\gamma(\theta)$ calculation}\label{tonks}

In the absence of a DDI, the dimensionless coupling parameter $\gamma$ due to the Van der Waals interaction is defined as~\cite{Olshanii:2001}
   \begin{equation}
       \gamma^\text{VdW} =  \frac{2}{n_\text{1D} |a_\text{1D}| } = \frac{m g^\text{VdW}_\text{1D}}{n_\text{1D} \hbar^2},
   \end{equation}
where $n_\text{1D}$ is the 1D particle density and  the interparticle interaction along the tube axis  is well approximated by an effective potential $U_\text{1D}=g^\text{VdW}_\text{1D}\delta(x)$.   The  1D Van der Waals  interaction strength is $g^\text{VdW}_\text{1D}= -2\hbar^2/(ma_\text{1D})$~\cite{Olshanii:1998jr}.  The effective 1D scattering length is given by 
   \begin{equation}\label{a1d}
   a_\text{1D} =  -\frac{l_\perp^2}{a_\text{3D}}  = -435(53)~\text{nm},
   \end{equation}
where $a_\text{3D}=141(17)$~Bohr is the weighted-average $s$-wave scattering length of $^{162}$Dy as measured in two previous experiments~\cite{Tang:2015pra,Tang:2016prl,Paule:1982} and  $l_\perp = \sqrt{\hbar/m\omega_\perp}=57.3(3)$~nm. 

With a DDI present,  $\gamma$  is 
  \begin{equation}
       \gamma(\theta) = \frac{m g^\text{total}_\text{1D}(\theta)}{n_\text{1D} \hbar^2},
   \end{equation}
where  $g^\text{total}_\text{1D}(\theta) = g^\text{DDI}_\text{1D}(\theta) +  g^\text{VdW}_\text{1D}$ and $ g^\text{DDI}_\text{1D}(\theta)$ is given in Eq.~\eqref{gddi1d}.   \hl{Confinement-induced resonances}  modify this expression for $a_\text{1D}$ through an additional factor of $\left( 1 - \frac{C a_\text{3D}}{\sqrt{2} l_\perp} \right)=0.87(2)$, where $C\approx 1.46$~\cite{Olshanii:1998jr,Haller:2010dj}. This correction does not significantly change the shape or magnitude of the theory curve in Fig.~\ref{Fig4}.   Moreover, this factor could be modified by the presence of the DDI to a value that has not been either measured or uniquely determined by theories of  \hl{ dipolar confinement-induced resonances} ~\cite{Sinha:2007gx,Bartolo:2013bj,Giannakeas:2013fd,Shi:2014ki,Guan:2014cc}.  Given this uncertainty, we choose to use the simple expression in Eq.~\eqref{a1d} for $a_\text{1D}$.

To find a weighed averaged $\gamma^\text{avg}(\theta)$, we calculate the number of atoms in each tube by assuming a Thomas-Fermi density distribution $n_{\mathrm{TF}}$ for the BEC:
   \begin{equation}
       n_{\mathrm{TF}} (\mathbf{r}) = \frac{15}{8\pi} \frac{N}{\prod_i{R_i}} \max{\left( 1-\sum_i \frac{r_i^2}{R_i^2}, 0 \right)},
   \end{equation}
where  $N=\int{ d \mathbf{r}^3 n_{\mathrm{TF}}}$   is the total atom number, $R_i$ is the Thomas-Fermi radius, and $i=x,y,z$. \hl{The TF approximation is justified given the weak dependence of $\gamma\sim N^{2/3}$ on atom number~\cite{Kinoshita:2006bg}. } We then obtain a 2D density distribution of the BEC in the $yz$-plane by integrating along the tube direction: 
    \begin{align}
       & n(y,z) = \int{ n_{\mathrm{TF}} (\mathbf{r}) dx} \\
              &= \frac{5}{2\pi} \frac{N}{R_y R_z} \left[\max{ \left( 1 - \frac{y^2}{R_y^2} - \frac{z^2}{R_z^2}, 0 \right) }\right]^{3/2}.
    \end{align}
To find the number of atoms loaded into each tube $N_{i,j}$, we assume each tube collects atoms in a square cross section with length $a=\lambda/2$, equal to the lattice site spacing, at a local density  $n(y,z)$ with the atom number given by  $N_{i,j} = a^2 n(y_i,z_j)$, where $y_i$ and $z_i$ denote the tube position. This calculation neglects rearrangements of atoms during the lattice loading procedure, i.e., tunneling when the lattice is still shallow, but this assumption is justified by the weak dependence of $\gamma$ on atom number.

We calculate the peak atomic density of each tube using the 1D Thomas-Fermi distribution  before the gas is excited.  Since $\gamma$ is only weakly dependent on atom number, we use the mean-field result rather than the full TG result, as in Ref.~\cite{Wenger:2016uc}. 
   \begin{equation}
       n^\text{TF}_{0} = \left[ \frac{9}{64} N^2_{i,j} \left( \frac{m \omega_x}{\hbar} \right)^2 |a_\text{1D}| \right]^{1/3}.
   \end{equation}
Before exciting the gas, each tube has a $\gamma^{i,j}_{0}(\theta)\propto 1/n^\text{TF}_{0}$, and for the ensemble of tubes, we calculate an average $\gamma^\text{avg}_{0}(\theta)$ weighed by atom number in each tube:
   \begin{equation}
       \gamma^\text{avg}_{0}(\theta) = \frac{ \sum_{i,j} \gamma^{i,j}_0 (\theta) N_{i,j} }{\sum_{i,j}{N_{i,j}}}.
       \label{gamma_initial_avg}
   \end{equation}
Note that for each tube, $\gamma^{i,j}_{0}(\theta)$ has a weak dependence on atom number: $\gamma^{i,j}_{0}(\theta) \sim N_{i,j}^{2/3}$.
   
For our experimental conditions, we load into approximately $70\times10$ tubes, with $\sim$50 atoms in the central tubes, resulting in an ensemble averaged density $n^\text{avg}_{0}  = 3.1$~$\mu$m$^{-1}$. This  yields an ensemble averaged initial $\gamma^\text{VdW,avg}_{0}=1.5(2)$.  Including the  $g^\text{DDI}_\text{1D}(\theta)$ term, $\gamma^\text{avg}_{0}(\theta)$ varies from 0.6(1) at $0^\circ$ to 1.9(2) at $90^\circ$.
   
The gas dephases at a time $\sim$100~ms after being diffracted. The dephasing reduces the density in each tube because the narrow, counterpropagating packets of atoms spread throughout the entire tube length with a higher classical turning point due to addition of the large energy from the momentum kick.  Our classical non-interacting dynamics simulation, discussed in Sec.~\ref{regime1}, shows that the dephased density distribution is approximately uniform, and we therefore estimate the dephased density to be $n_\text{d}^{i,j} = N_{i,j}/(2d_\text{m})$, where $d_\text{m}=\hbar k_\text{D}/(m \omega_x)=12$~$\mu$m is the maximum distance an atom travels away from the trap center. The dephased ensemble-averaged $\gamma^\text{avg}_{d} (\theta)$ is then found by replacing $\gamma^{i,j}_{0}$ in Eq.~\eqref{gamma_initial_avg} with $\gamma^{i,j}_{d} \propto 1/n_\text{d}^{i,j}$. For the aforementioned experimental parameters, we find $\gamma^\text{VdW,avg}_\text{d}=5.7(7)$, with an ensemble-averaged dephased density $n^\text{avg}_\text{d}=0.8$~$\mu$m$^{-1}$.  Including the  $g^\text{DDI}_\text{1D}(\theta)$ term, $\gamma(\theta)\equiv\gamma^\text{avg}_{d}(\theta)$ varies from 2.2(3) at $0^\circ$ to 7.4(9) at $90^\circ$, as shown in Fig.~\ref{Fig1}(c).

In its ground state, a system at such values of $\gamma$ would be \hl{in the crossover to the} TG regime, in which the microscopic bosons exhibit antibunching, as free fermions would~\cite{Kinoshita:2004jp,Kinoshita:2005iz}. This antibunching occurs because the interaction \hl{strength} dominates the zero-point energy.  Whether fermionization persists in the high-energy, far-from-equilibrium post-kick evolution is a priori unclear, though in equilibrium, at the post-kick energy density, there is no antibunching~\cite{Kheruntsyan:2003dk}. Since the dephased state is far from equilibrium, these results cannot be applied directly, but are suggestive (as one might expect typical high-energy-density states to be thermal in some respects~\cite{Rigol:2009ew}).
   
\begin{figure}[t!]
\includegraphics[width=1\columnwidth]{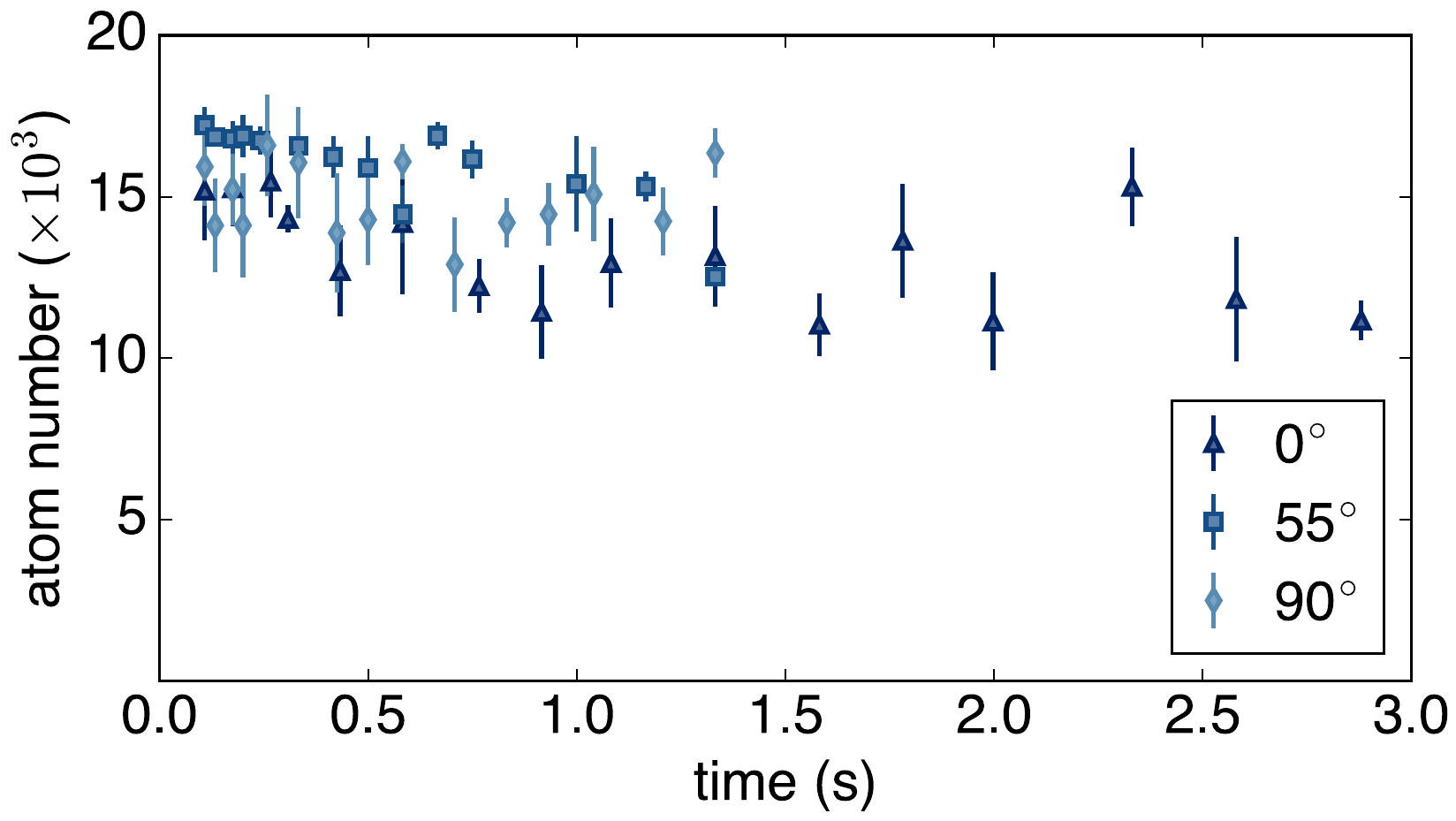}
\caption{Atom number during the momentum evolution at $\gamma^\text{VdW,avg}_{d}=5.7(7)$ for $\theta = 0^{\circ}$ (triangle), $55^{\circ}$ (square), and $90^{\circ}$ (diamond). }
\label{fig:number_vs_time}
\end{figure}

\section{Atom number variation}\label{atomnumber} 

Any atoms that flip spin due to spontaneous emission from the optical trap and lattice beams are immediately lost from dipolar relaxation collisions and do not lead to heating of the gas~\cite{Burdick:2015bx}. Feshbach resonances are avoided by tuning to 1.58(1)~G, which lies within a resonance-free region between 0.5~G and 2.5~G~\cite{Baumann:2014ey}. We ensure that the B-field remains at this field within 10~mG at every angle.  We do not observe any confinement-induced resonances or dipolar confinement-induced resonances since we do not observe resonant atom loss at any $\theta$ angle investigated~\cite{Ronen:2006dw,Sinha:2007gx,Haller:2009jrb,Haller:2010dj,Giannakeas:2013fd,Bartolo:2013bj,Guan:2014cc,Shi:2014ki}.

We do not observe significant atom loss in the data sets presented. Atom number as a function of time is shown in  Fig.~\ref{fig:number_vs_time} for $\theta = 0^{\circ}$, $55^{\circ}$, and $90^{\circ}$, with $\gamma^\text{VdW,avg}_{d}=5.7(7)$. For the longest observation time of 2.8~s, we lose about 25\% of the total atoms, which increases  $\gamma^{i,j}_{0}$ by just 16\%  according to the $\gamma_{i,j} \sim N_{i,j}^{2/3}$ scaling relation. Aside from atom number loss during the observation time, there is also a slight variation of atom numbers between data taken for different $\theta$. For all angles used in the $\gamma^\text{VdW,avg}_{d}=5.7(7)$ measurement, the mean atom number is $15(2)\times 10^3$, corresponding to a 13\% variation, which is smaller than the 25\% variation in atom number over the time evolution at a fixed $\theta$. We therefore conclude that it is reasonable to treat $\gamma^\text{VdW,avg}_{d}$ as constant in interpreting our data for different $\theta$; i.e., the observed trend in thermalization time cannot be explained by variation in atom number.  This constancy of $\gamma^\text{VdW,avg}_{d}=5.7(7)$ versus time is in contrast to the rapid increase in $\gamma$ observed in Ref.~\cite{Kinoshita:2006bg} due to large atom loss rates. 
   
\begin{figure}[t!]
\includegraphics[width=1\columnwidth]{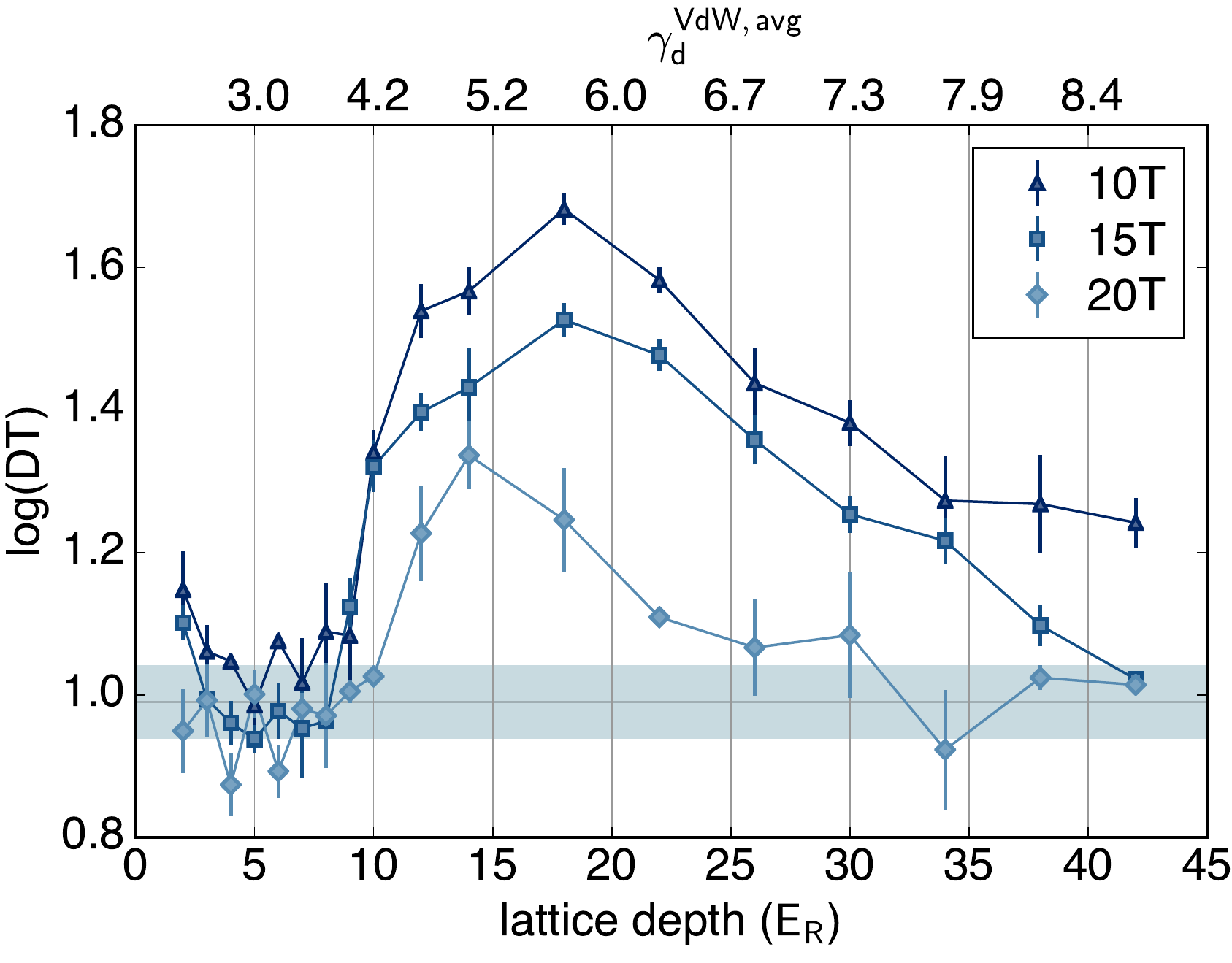}
\caption{Thermalization distance at $\theta=90^{\circ}$ measured as a function of lattice depth  for different observation times: 10$T$ (triangle), 15$T$ (square), and 20$T$ (diamond). The blue band represents the noise floor and its $1\sigma$ uncertainty. The  $\gamma^\text{VdW,avg}_\text{d}$'s associated with the lattice depths are shown on top.}
\label{fig:lattice_depth_scan}
\end{figure}

\section{Heating measurements and simulations}\label{heating}

Heating from the lattice beams can affect the momentum distribution evolution. The lattice lasers can induce heating in two ways: 1) Intensity noise at certain frequencies can parametrically heat the gas or excite atoms to higher lattice bands; 2) Spontaneous emission imparts photon recoil momentum onto the atoms, whose projection along the tube direction leads to heating~\cite{Riou:2012ia,Riou:2014gd}. We now show that the second mechanism is the dominant heating source in our system before discussing its effect on the DT in more detail with the aid of a collisionless Monte Carlo simulation.

We measure the heating rate in our system by loading a BEC into the lattice and measuring its momentum distribution versus lattice hold time. The procedure is identical to the DT measurements, though without splitting the gas. At short hold time ($t<0.5$~s at $18E_\text{R}$), we observe a  distribution along $\hat{x}$ that is similar to that reported in Ref.~\cite{Kinoshita:2006bg}: a broad Gaussian centered about a narrower Gaussian. At longer times, the measured distribution fits well to a single Gaussian. The fitted width of the single Gaussian increases linearly with time, and the best-fit slope corresponds to the heating rate. We verify that the dominant heating mechanism in our system is spontaneous emission by observing that the heating rate of an unkicked gas decreases as $1/\Delta$ when we vary the detuning $\Delta$ from atomic resonance at constant lattice depth $V_0$.

Engineering a TG system requires the deepest lattice possible. However, too much heating from a large $V_0$ would obscure the dynamics of interest. We therefore search for a $V_0$ with the slowest thermalization rate. We experimentally determine this optimal depth by measuring the DT at a fixed holding time for a range of $V_0$ values at $\theta=90^{\circ}$, the angle with the largest DDI. The results for three different holding times, $t=10T$, $15T$, and $20T$, are shown in  Fig.~\ref{fig:lattice_depth_scan}. The slowest thermalization occurs near $V_0=18E_\text{R}$, which is the lattice depth we use for our measurements, yielding $\gamma^\text{VdW,avg}_\text{d}=5.7(7)$.

\begin{figure}[t!]
\includegraphics[width=1\columnwidth]{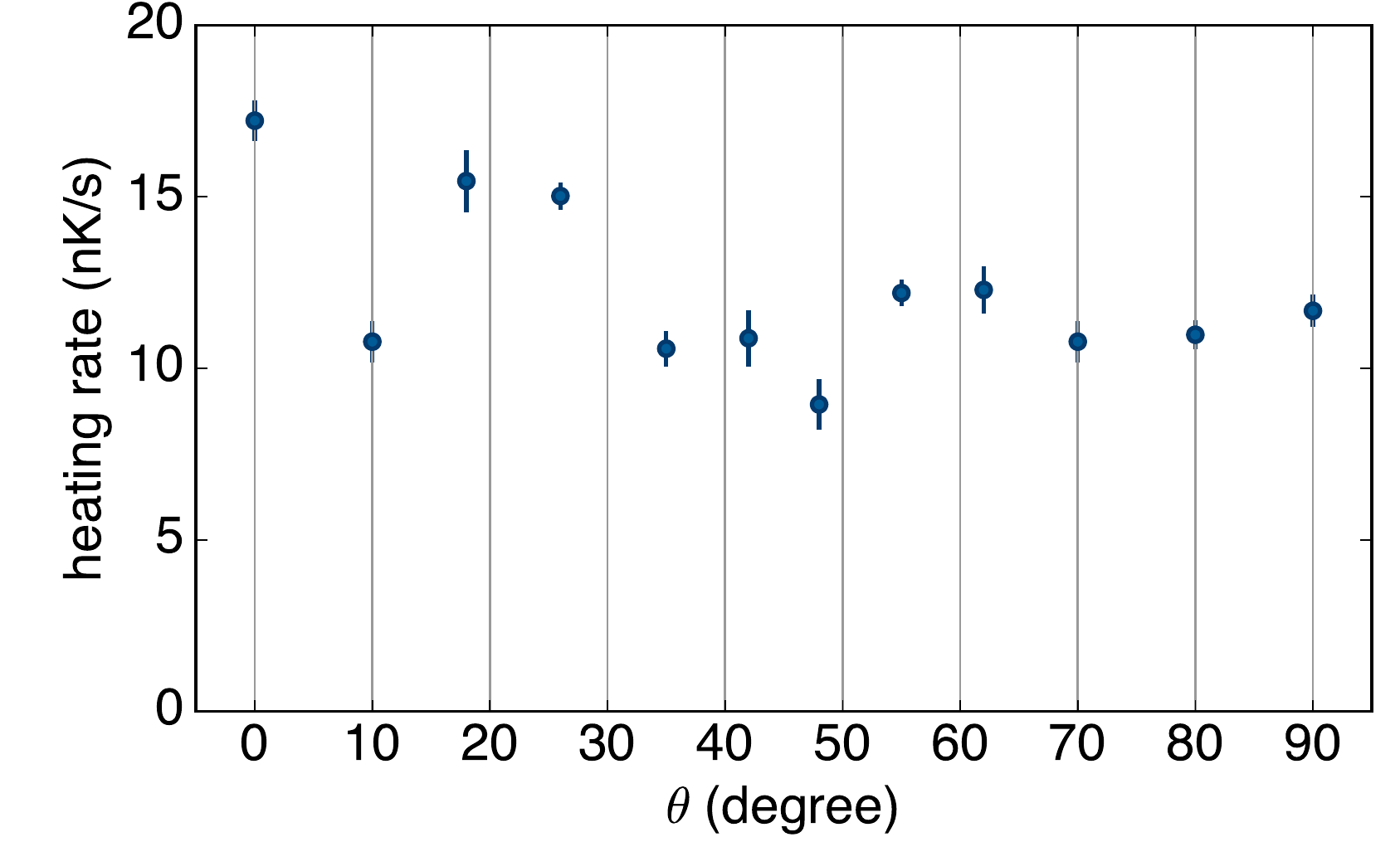}
\caption{Measured heating rate of the undiffracted BEC at equilibrium in a $V_0=18E_\text{R}$ 2D lattice at various angles.}
\label{fig:heating_vs_angle}
\end{figure}

We measured the heating rate at $V_0=18E_\text{R}$ for the twelve $\theta$ values used in our thermalization rate data. The results are shown in Fig.~\ref{fig:heating_vs_angle}. The heating rates are similar among all angles with little-to-no systematic variation.  The highest rate, 17(1)~nK/s at $0^{\circ}$,  is still $\sim$5$\times$ slower than the slowest rate observed in Ref.~\cite{Kinoshita:2006bg}. The low heating rates versus those in Ref.~\cite{Kinoshita:2006bg} are achieved through the use of lower 2D lattice depths and 5--10$\times$ smaller $n_\text{1D}$.  Nevertheless, in Ref.~\cite{Kinoshita:2006bg} the ratio between collision energy and transverse trap frequency is $E_c/(\hbar \omega_{\perp})=0.45$, whereas we have 0.47---essentially the same. However, the recoil momentum of the lattice $k_\text{R}$ used in Ref.~\cite{Kinoshita:2006bg} is $\sqrt{2}\times$ larger than ours. In addition, the mass of their atomic species, Rb, is twice lighter than Dy's. Therefore, their $\omega_{\perp}$ has to be four times larger than in our experiment, and so a much deeper lattice is required to remain in the 1D regime, leading to larger heating rates. On the other hand, our shallower lattice results in a faster intertube tunneling rate $J$. However, we estimate in Sec.~\ref{tunneling} that the thermalization rate associated with tunneling is $\geq$10$\times$ smaller than our lowest measured thermalization rate and is therefore negligible. 

We use the measured heating rate of a BEC at equilibrium to simulate the effect of spontaneous emission heating on the DT evolution of an experimentally measured dephased distribution. A Monte Carlo method is employed that accounts for the heating effects described in Refs.~\cite{Riou:2012ia,Riou:2014gd}, and we find that the dominant heating process is from one-body spontaneous emission.   Following Ref.~\cite{Riou:2012ia}, we consider those changes in vibrational state in the transverse direction with $n = n_y + n_z \rightarrow n\pm 1$ due to both absorption and emission of lattice photons. Atoms with $n\ge 3$ are considered lost from the trap, since we expect intertube tunneling for atoms in these states to become non-negligible because their vibrational energy approaches the transverse lattice depth. Atom loss can also occur in the axial direction when the total axial energy for an atom exceeds the axial trap depth $V_0 = m\omega^2 w_0^2/4$, where $\omega$ and $w_0$ are the trap frequency and Gaussian beam waist in the direction of interest, respectively. As in the experiment, we observe little atom loss in the simulation: The typical loss is 3\% in 8~s, which is over two times longer than the longest thermalization time measured in the experiment.

In addition to one-body loss due to spontaneous emission, two-body collisions after a spontaneous emission event can lead to heating. In particular, Ref.~\cite{Riou:2014gd} considers seven two-body transverse state-changing collisional processes that are energetically allowed and permitted by parity selection rules. As the vibrational levels of the scattered atoms are modified, there is a finite probability for the transverse energy to be deposited in the axial direction, leading to an axial momentum kick. Since the rates of such transitions depend on the population of the relevant $n\neq 0$ states, these collisions are second-order; the atoms are initialized in the $(n_y,n_z) = (0,0)$ state and spontaneous emission is the only mechanism to excite them to higher vibrational levels. Indeed, by using the worst-case reflection and transmission probabilities~\cite{Berges:2004}, we find within the experimental timescale that the simulated momentum distributions exhibit negligible deviation compared to those without two-body collisions.  Therefore, we need only consider one-body heating processes in our analysis. This is fortunate, as dipolar two-body collisions could have led to $\theta$-dependent heating. 

We can now use these heating rates as inputs to simulations of the DT evolution. This is done in order to account for how heating affects the rate of change of DT so that we may then account for heating in our measured thermalization rates.  To do so, we introduce to the simulated momentum distributions the measured heating rate and a Gaussian white noise background that is matched to the experimentally measured noise level. The DT and noise floor are then computed in the same way as described in Sec.~\ref{DTmetric}.  To reduce Monte Carlo sampling noise, we average twenty simulated distributions so that their noise is negligible compared to the added detection noise.  The simulations yield thermalization rates between $0.156(5)$~s$^{-1}$ and $0.25(1)$~s$^{-1}$ versus $\theta$.  This shows that the fastest spontaneous-emission-limited heating rate is slower than the slowest measured thermalization rate, and therefore this heating rate is never larger than our measured thermalization rates, and indeed is much smaller than those rates for $\theta$'s above $30^\circ$.  Finally, to deduce the thermalization rate due to the integrability-breaking physics alone, i.e., the rates plotted in Fig.~\ref{Fig4}, we subtract the simulated spontaneous emission-induced thermalization rates from the experimental thermalization rates.

\section{Comments on DT metric}\label{dataanalysis}

We have  considered other metrics for distance-to-thermalization (DT): (1) Kurtosis is a common measure of deviation from a Gaussian for a given distribution, but does not work well for our data due to its high sensitivity to noise. 2) We compared $p(x)$ to a thermal distribution with the same total energy at each recorded time step, including both initial kinetic energy imparted on each atom and heating from spontaneous emission. This method is less reliable because the energy summation from $p(x)$ is sensitive to noise in the high-momentum wings. 

\section{Measurement of DT for different $\theta$ and rotation times}\label{rotation}

 \begin{figure}[t!]
\includegraphics[width=1\columnwidth]{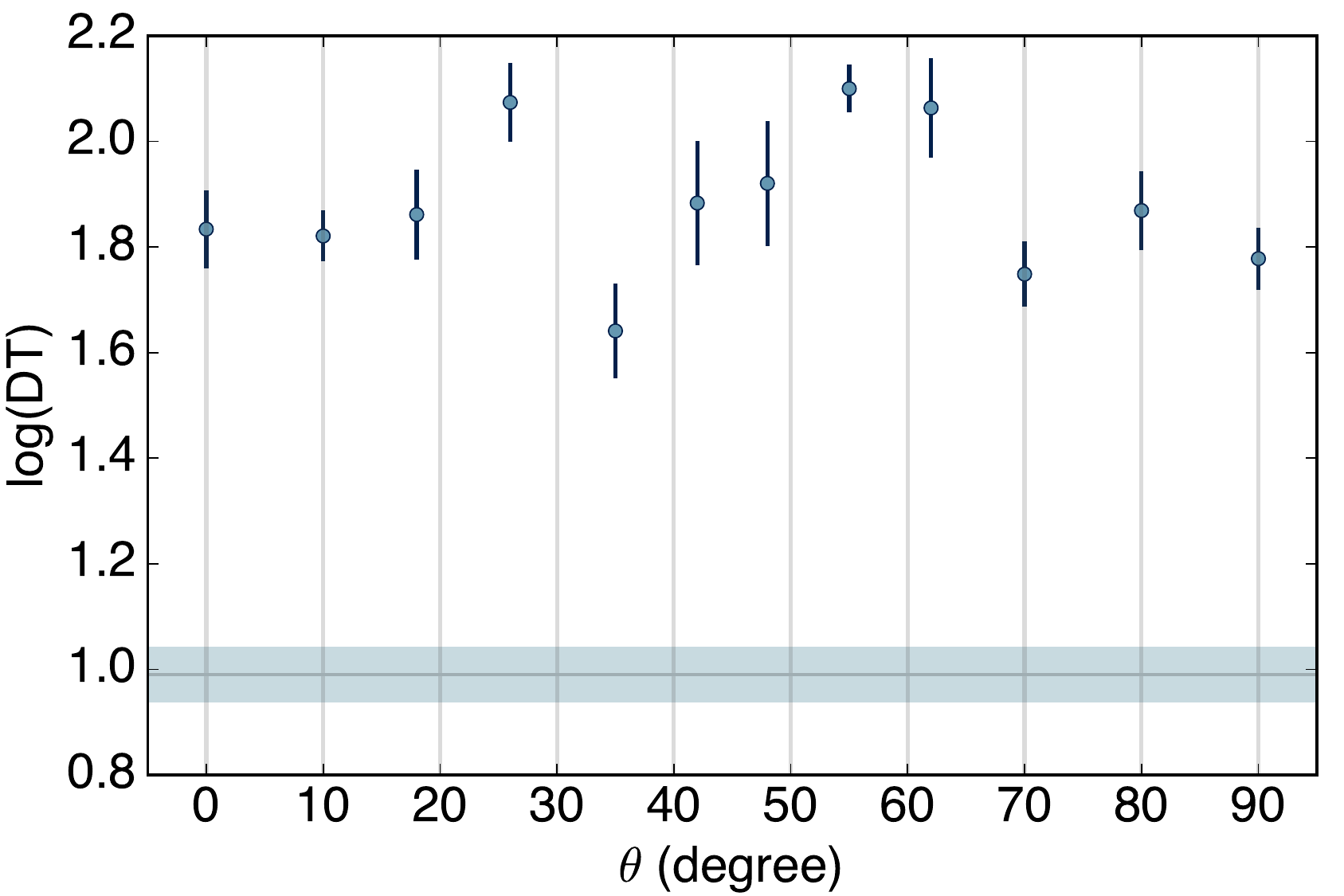}
\caption{Distance-to-thermalization (DT) of the field-rotated state near the end of regime I for each $\theta$ value investigated. The blue horizontal line is the mean noise floor, and the light blue band represents its $1\sigma$ uncertainty.}
\label{fig:initial_state}
\end{figure}

\hl{The kicking procedure described in Sec.~\ref{kicking} ensures that the boundary between regime I and II appears at $\log(\text{DT})\approx1.5$ regardless of the final $\theta$ setting. We recall that this is because, to eliminate systematics due to the splitting processes, we fix the system to evolve under $\theta=35^\circ$ in regime I before rotating $\theta$ to its final setting at the beginning of regime II.  Figure~\ref{fig:initial_state} shows data exhibiting no systematic variation in DT versus $\theta$ at the end of regime I. }

\begin{figure}[t!]
\includegraphics[width=1\columnwidth]{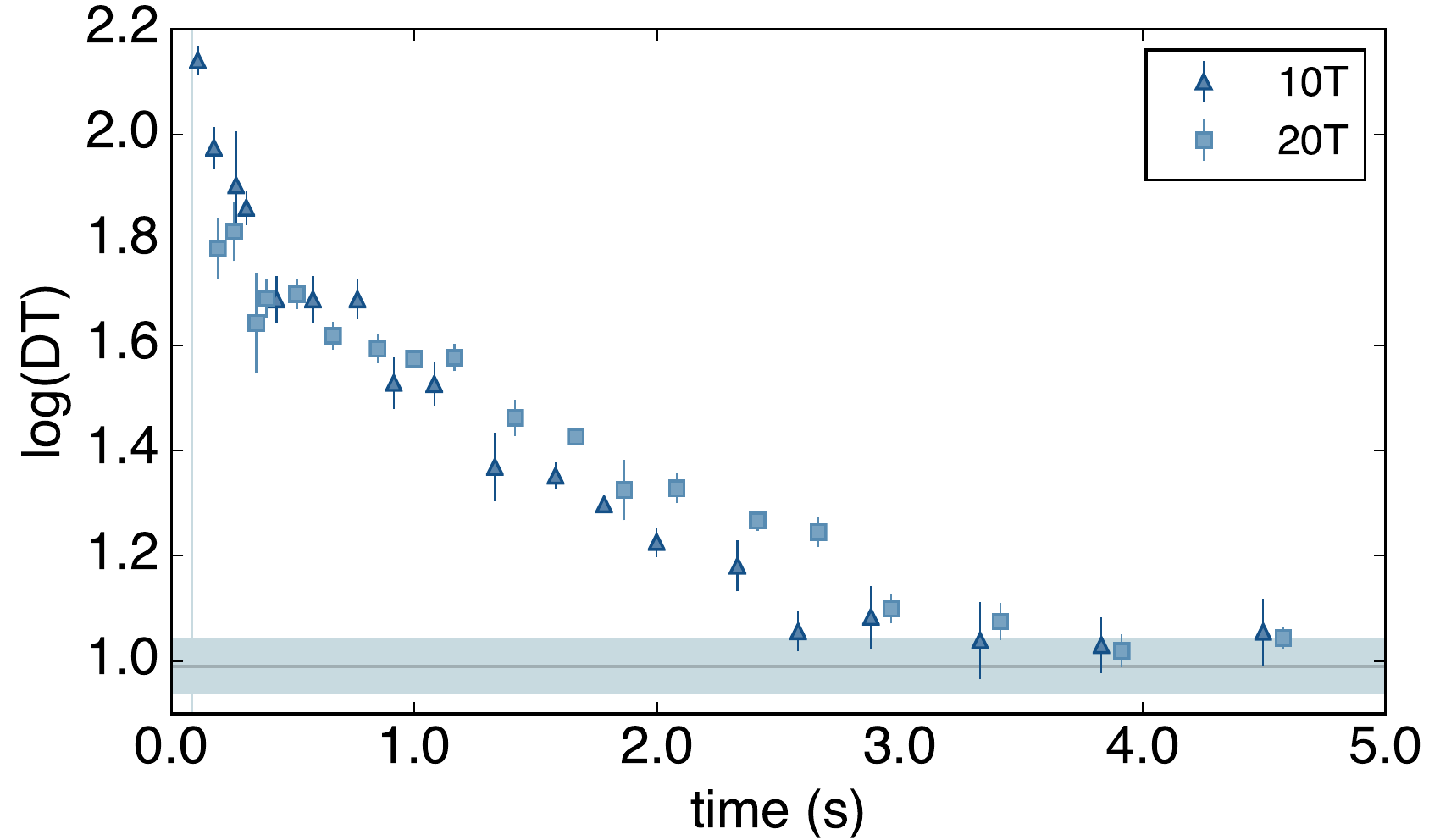}
\caption{Time evolution of DT at $\theta=0^{\circ}$ for rotating the field after a waiting time of $10T$ (triangle) and $20T$ (square).}
\label{fig:waiting_time_comparison}
\end{figure}

While for some $\theta$ there can be an additional dephasing evolution after the rotation time---e.g., the $0^\circ$ data in the inset of Fig.~\ref{Fig3} takes longer to reach regime II ($\log(\text{DT})\approx1.5$) than the $90^\circ$ data---we have verified that waiting longer to rotate does not affect the subsequent thermalization rate: We took data at two different rotation times  for $\theta = 0^\circ$ to demonstrate that the time chosen for the rotation also does not affect the subsequent thermalization rate.  We choose $\theta = 0^\circ$ because it exhibits the slowest decay time so that we can best test the difference between these rotation times.  These data are shown in  Fig.~\ref{fig:waiting_time_comparison}, and we find that the slow decay rate is approximately unchanged within experimental resolution. 

\section{Tunneling between tubes}\label{tunneling} The tunneling rate is approximately given by
        \begin{equation}
            \frac{J}{E_R} \simeq \frac{4}{\sqrt{\pi}} s^{3/4} \exp(-2 s^{1/2}),
        \end{equation}
where $s=V_0/E_R$ is the dimensionless lattice depth. This formula agrees with the exact value of $J$ to better than 10\% accuracy for $s>15$~\cite{Bloch:2008gl}. For our lattice, $s=18$ and $J=0.004E_R$, which corresponds to a $J/\hbar=2\pi \times9$~Hz tunneling rate. 
    
Tunneling between tubes can break integrability by allowing effectively 2D scattering~\cite{Yurovsky:2008ft}: To leading order, two particles in the same tube collide while one atom scatters to a neighboring tube via tunneling. Such a scattering event conserves total momentum and energy, but not momentum along the tubes due to a finite lattice bandwidth $4J$, leading to thermalization.
    
We estimate the thermalization timescale set by tunneling in the following manner. The initial momentum $k_i=2k_\text{D}$ and final momentum $k_f$ of an atom in a two-particle scattering event that involves tunneling  can be related by     
    \begin{equation}
        \frac{(\hbar k_i)^2}{2m} - \frac{(\hbar k_f)^2}{2m} = 4J,
    \end{equation}
from which we obtain the relative change in momentum $\Delta k = k_i - k_f \approx (4J/2E_\text{R}) k_\text{D}$. Such scattering events involving tunneling then lead to a random walk in momentum space with a step size of $\Delta k$.  Thermalization requires a change in momentum on the order of $k_i$, and the time $T_\text{th}$ that it takes for this process is given by
    \begin{equation}
        \Delta k\sqrt{T_\text{th} (2f_l) RN/2}  = k_i,
    \end{equation}
where  the collision rate is twice the longitudinal trap frequency $f_l=T^{-1}=\omega_x/(2\pi)$ and the square root on the left side arises from the random walk process.  The factor of $RN/2$ is the total number of collisions given $N$ atoms in each tube with reflection coefficient $R = (2k_\text{D}a_\text{1D})^{-2}=1/28$ per atom~\cite{Kinoshita:2006bg}.  For our parameters, $T_\text{th} \approx600$~s. This estimated time scale is two orders of magnitude larger than the longest measured thermalization time. We also note that tunneling cannot be the source of the angular dependence  we observe in the data. Therefore, we conclude that tunneling between tubes, while not completely negligible, is a much smaller thermalization mechanism than either the DDI or the spontaneous heating caused by the lattice lasers.

\section{Collisionless classical dynamics simulation}\label{collisionless}

As mentioned in Sec.~\ref{regime1}, there are two dephasing processes due to the trap:  (1) dephasing of oscillations between different harmonic tubes due to their different natural frequencies; and (2) dephasing of the oscillations of the gas in a single tube due to its anharmonicity. To quantify their relative contributions to dephasing, we simulate the momentum distribution evolution using a classical dynamics model that does not take account for interactions.

\hl{We numerically solve the classical equation of motion for atoms in each tube given its longitudinal Gaussian potential. The parameters of the Gaussian potential of each tube are determined by the tube location with respect to the center of the crossed ODT trap. We simulate an array of 70$\times$12 tubes in the $yz$-plane. The ODT beam parameters are given in Sec.~\ref{BEC}. We assume the foci of the two crossed ODT beams overlap perfectly and neglect the axial trapping contribution from the lattice beams. We note that any slight imperfection in beam alignment and beam shape distortion only increases the effect of (1) and (2); consequently, the simulation provides an upper bound to the dephasing time.}

\hl{We initialize the simulation by distributing $15\times10^3$ atoms into the tubes using the density calculation described in Appendix~\ref{tonks}. All atoms are located  in the center of each tube and to match the measured initial momentum spread, we assign an initial momentum to each atom by sampling from a Gaussian distribution with standard deviation $\sigma=0.2 \hbar k_\text{D}$. We then add a $\pm 2\hbar k_\text{D}$ momentum kick to the atoms in each tube.  We solve the trajectory for each atom at a time step of $T/30$, where $T=16$~ms is the trap period of the central tube. Doing so allows us to keep track of the momentum and position of each atom at every time step.}

\begin{figure}[t!]
    \includegraphics[width=1\columnwidth]{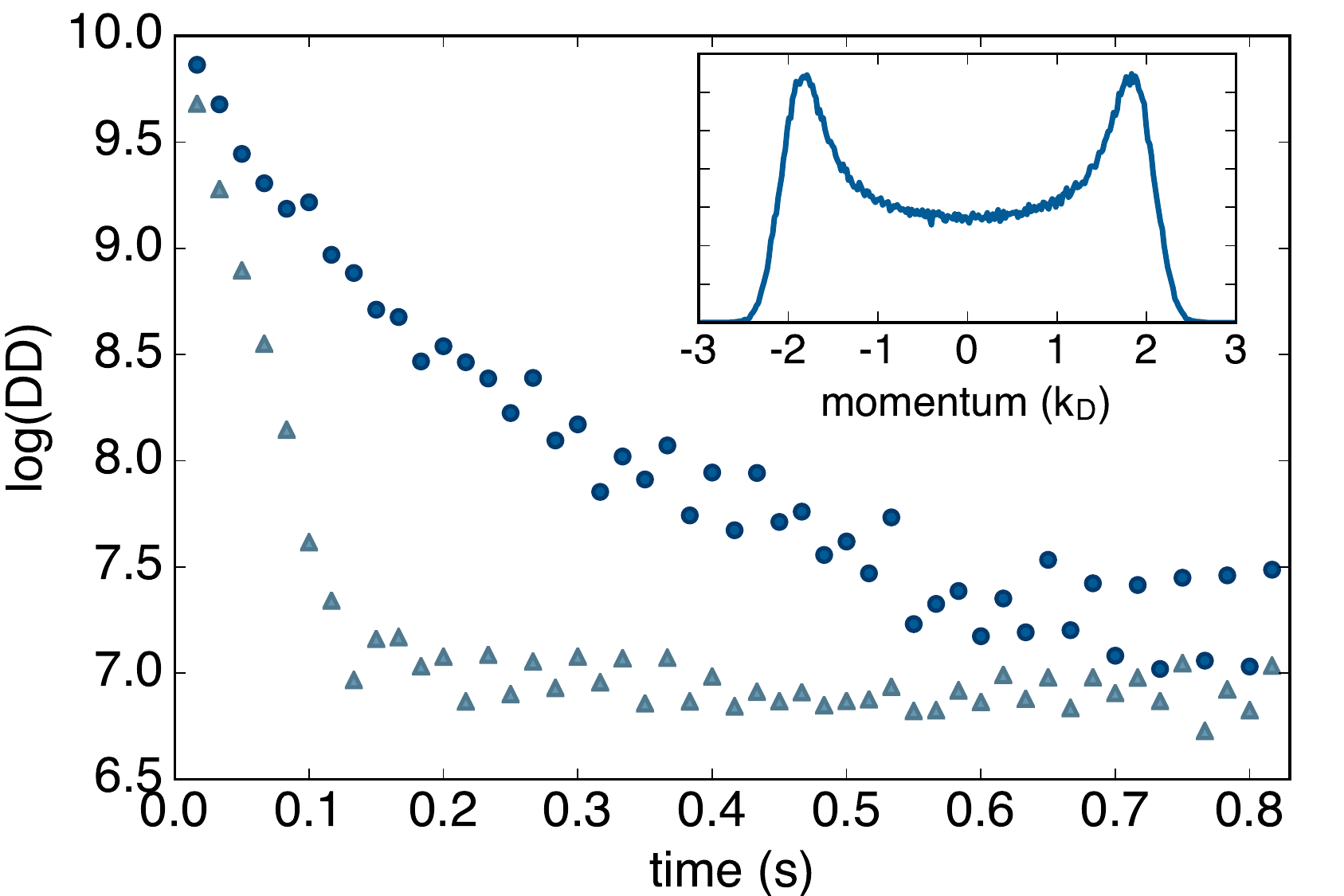}
    \caption{\hl{Time evolution of distance--to--dephasing (DD) predicted by the collisionless classical dynamics simulation for two cases: the first for when only process (2) is present (circle), and second when both (1) and (2) are present (triangle). Inset shows the dephased momentum distribution.}}
    \label{fig:classical_dephasing}
    \end{figure}

\hl{Unlike in the experiment, the momentum distribution produced by this collisionless simulation is not stationary within a period. However, the averaged distribution over a period reaches a steady state, as shown in the inset of Fig.~\ref{fig:classical_dephasing}. The figure's main panel shows the RMS difference between the averaged distribution over each period and this steady distribution.  We call this metric the distance--to--dephasing (DD).  This metric is more appropriate than DT because we are interested in the time to the end of technical dephasing.  The figure shows  two simulation cases:  the first (shown as light blue triangles) only includes process (2); the second (shown as darker blue circles) includes both processes (1) and (2).  We see in Fig.~\ref{fig:classical_dephasing} that if only process (2) is present, the momentum distribution completely dephases in about 600~ms, while  the dephasing time reduces to  $\sim$150~ms when process (1) is included.  Both processes are important, and the non-negligible contribution of process (2) means it is reasonable to assume an equilibrated density profile when calculating dipolar interactions and $\gamma$.} 

\section{Details of single-sided kick measurement}\label{dephasing}

\begin{figure}[t!]
\includegraphics[width=1\columnwidth]{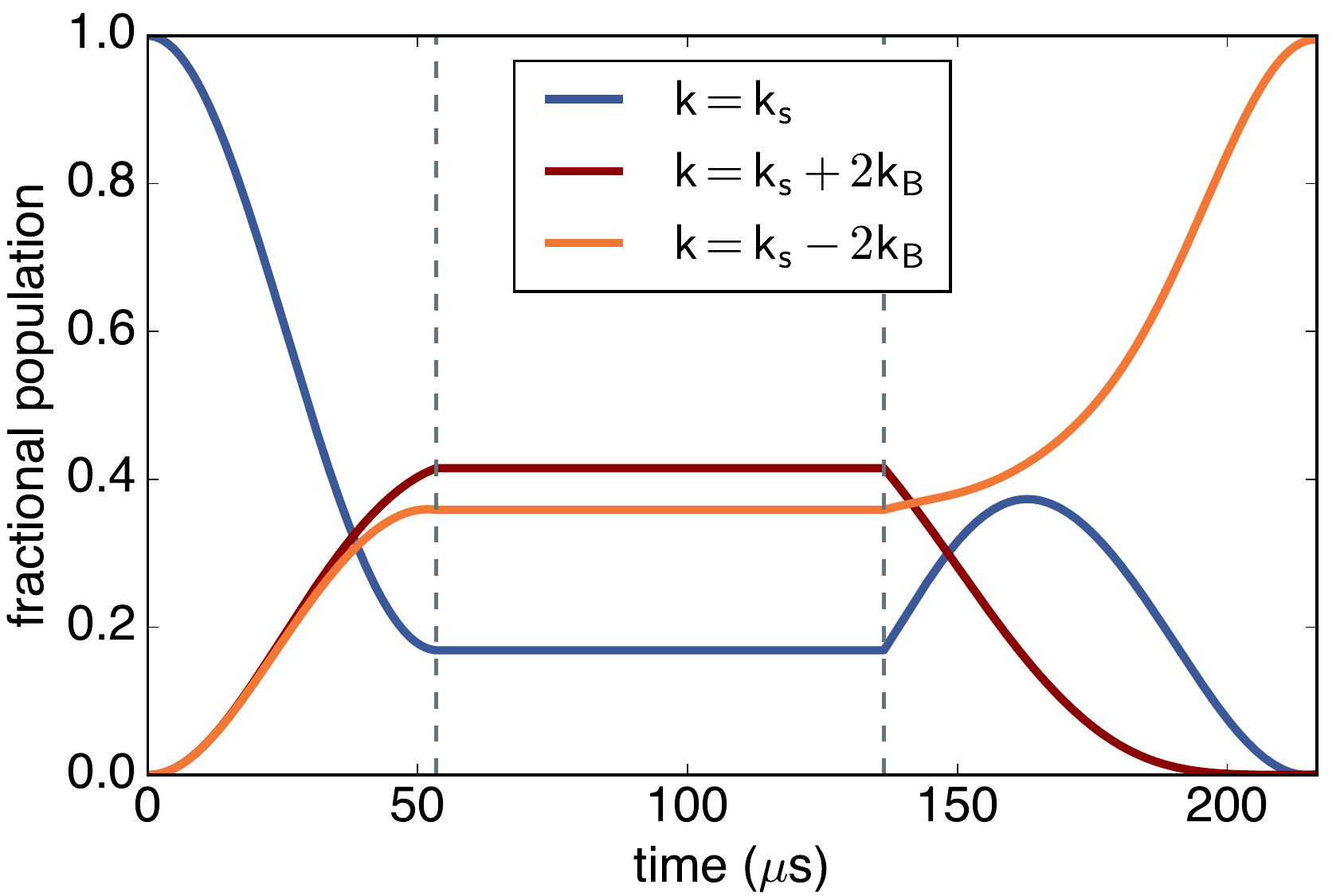}
\caption{Calculated time evolution of the fractional populations of the $|\hbar (k_s \pm 2 k_\text{D})\rangle$ states and the undiffracted state $|\hbar k_s\rangle$ during the single-sided kick pulse sequence. See text for detailed settings for each parameter.}
\label{fig:single_kick_population}
\end{figure}

We impart a single-sided momentum kick $p \approx -2 \hbar k_\text{D}$ to the atoms while keeping all other trap settings identical to that employed to take the thermalization data in Fig.~\ref{Fig3}.  We then measure the time at which the resulting distribution dephases. The single-sided kick is achieved using a double-pulse sequence similar to that used for creating the symmetric $|\pm 2 \hbar k_\text{D}\rangle$ splitting. The spatial symmetry is broken by introducing a small initial momentum 
$k_s$ to the BEC. Using a numerical optimization algorithm, we find that nearly all atoms can be transferred to the $|k_s - 2 \hbar k_\text{D}\rangle$ state using the following parameters: $k_s=-0.21 k_\text{D}$, a phase grating
lattice depth of $11.1 E_r/2$, a first pulse with duration $\tau_1 = 60~\mu$s, followed by $\tau_2 = 93~\mu$s of free-evolution, and a second pulse with duration $\tau_3 = 90~\mu$s. The calculated time evolution of the populations of the lowest two diffraction orders and the undiffracted order are shown in Fig.~\ref{fig:single_kick_population}. Populations in the higher diffraction orders are negligible.

To compare the single-sided dephasing time to our thermalization data, \hl{and to facilitate the pinpointing of the dephasing time,} we add to the single-sided distribution its own mirror image to emulate the situation where there are two packets of atom oscillating symmetrically in the tube. We then use the same analysis procedure as described earlier to find the $DT(t)$. The results are shown in Fig.~\ref{fig:dephasing_dist}.

\begin{figure}[t!]
    \includegraphics[width=0.6\columnwidth]{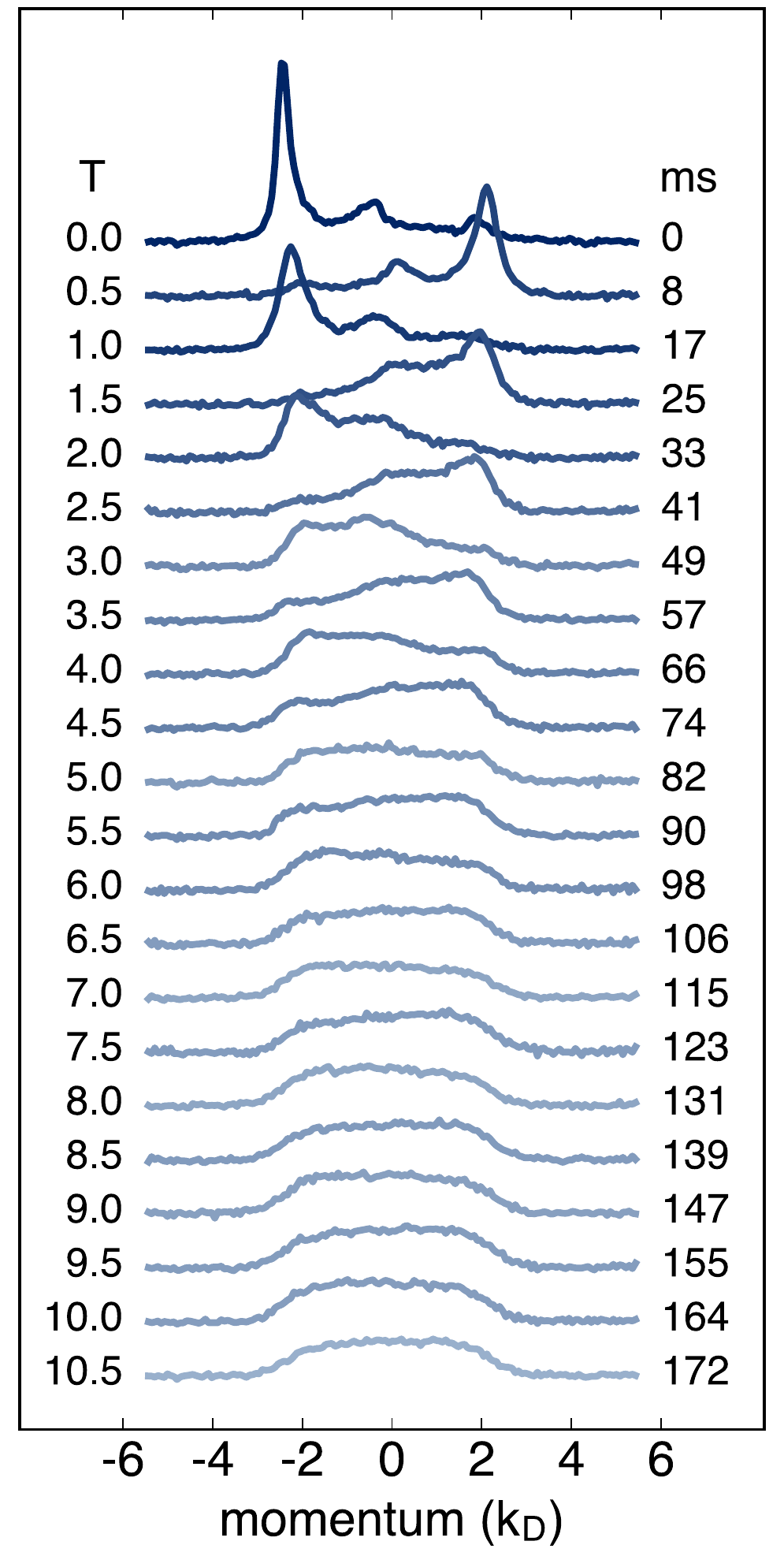}
    \caption{Time evolution of an initially single-sided momentum distribution, whose dephasing time scale accounts for the \hl{technical dephasing mechanisms while minimizing those from high-energy collisions.}}
    \label{fig:dephasing_waterfall}
    \end{figure}

\begin{figure}[t!]
    \includegraphics[width=1\columnwidth]{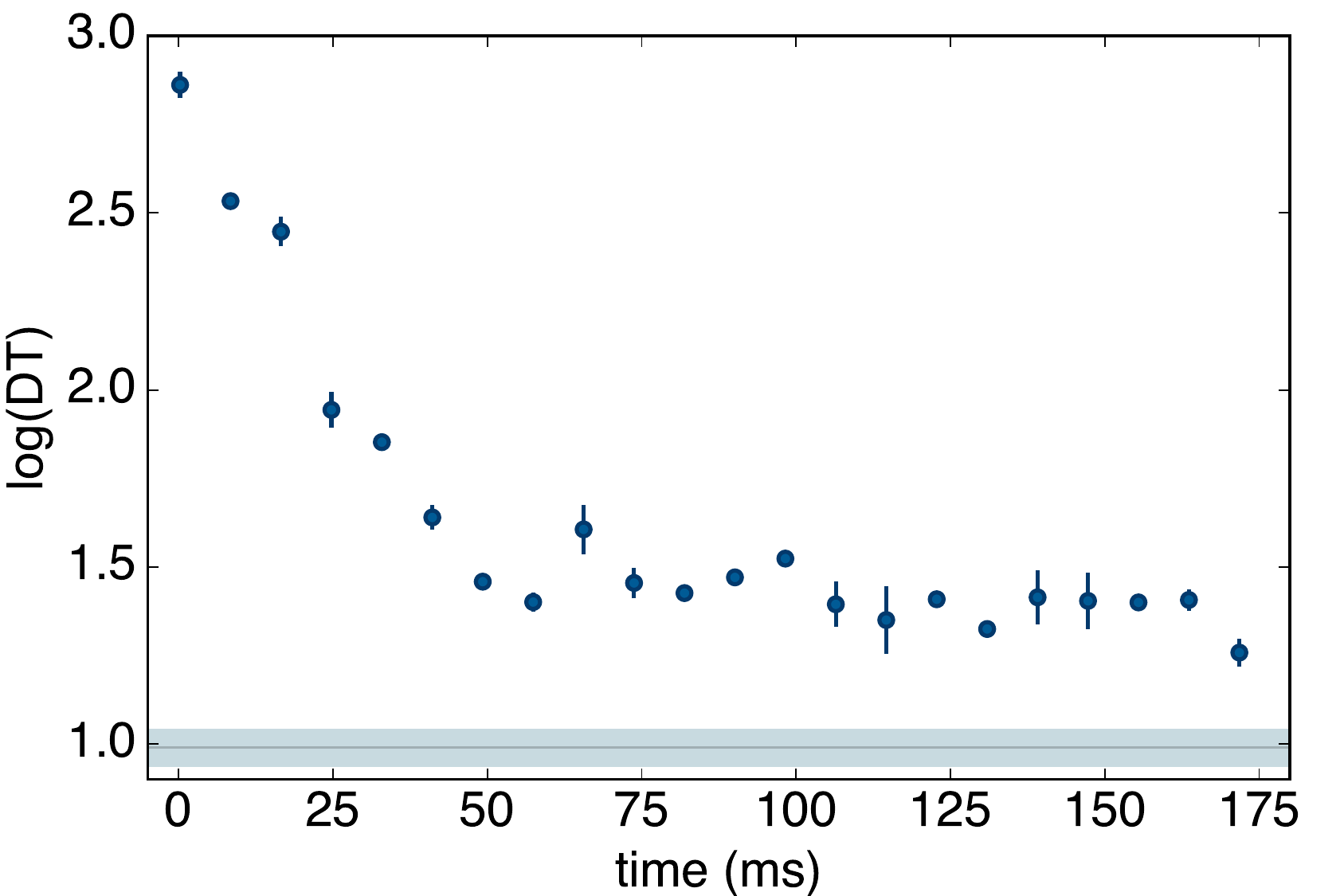}
    \caption{Thermalization distance of synthesized symmetric distributions generated by adding to the single-sided distributions shown in Fig.~\ref{fig:dephasing_waterfall} their own mirror images. The blue horizontal line is the mean noise floor, and the light blue band represents its $1\sigma$ uncertainty.}
    \label{fig:dephasing_dist}
    \end{figure}

The observed single-sided momentum distribution evolution is shown in  Fig.~\ref{fig:dephasing_waterfall}. We note that the value of DT at this dephasing time, $\log{(\text{DT})}\approx1.5$, is consistent with the choice of division between regime I and II in the thermalization data of Fig.~\ref{Fig3}.

The experimental sequence ensures that the atoms experience the same level of anharmonicity and inhomogeneity as in the measurements starting with a symmetric $|{\pm} 2 \hbar k_\text{D}\rangle$ distribution, but removes the effects of high-momentum interactions, i.e., head-on collisions.  In this configuration, the collisions alone (in the absence of anharmonicity) \textit{cannot} give rise to a stationary momentum distribution, and so the stationary momentum distribution must arise from the technical dephasing processes. (The generalized Kohn theorem~\cite{brey89} guarantees that the oscillatory center of mass motion in a strictly harmonic trap---and hence the oscillations of the momentum distribution---are unaffected by interactions.)

\section{Exact diagonalization calculations}\label{XXZcalc}

\subsection{Momentum distribution functions}\label{XXZcalc1}

\begin{figure}[t!]
    \includegraphics[width=1\columnwidth]{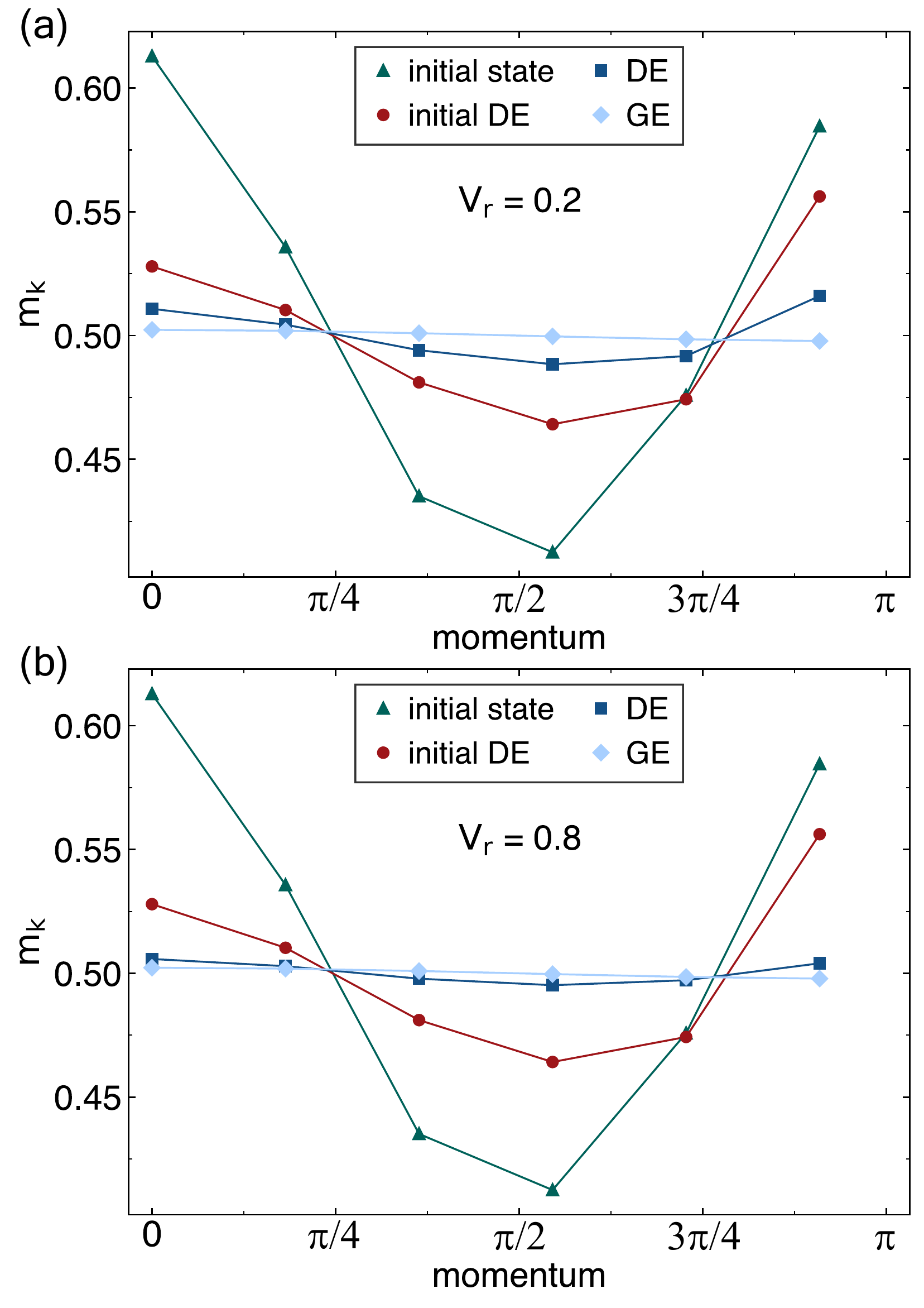}
    \caption{Examples of momentum distributions ($m_k$) obtained in the exact diagonalization calculations of the two-rung  hard-core boson model with $L=22$. We show the momentum distributions for an initial state (labeled as ``initial state''), the diagonal ensemble after the quench to the integrable part of $\hat H_F$ (labeled as ``initial DE''), the diagonal ensemble after the quench to $\hat H_F$ (labeled as ``DE''), and the grand canonical ensemble prediction for the thermal momentum distribution after the quench (labeled as ``GE''). For the quenches, we set $V = 1.6$ and (a) $V_r = 0.2$ and (b) $V_r = 0.8$.  The momentum distribution of the initial state and of the diagonal ensemble after the quench to the integrable part of $\hat H_F$ are the same in (a) and (b), only the diagonal and grand canonical ensemble predictions change due to the change of $V_r$.}
    \label{fig:XXZmomentum}
\end{figure}

In Fig.~\ref{fig:XXZmomentum}, we show examples of momentum distribution functions of: (i) initial states, (ii) the diagonal ensemble (DE) after the quench to the integrable part of $\hat H_F$, (iii) the DE after the quench to $\hat H_F$, and (iv) the grand canonical ensemble (GE) prediction for the thermal momentum distribution after the quench (see Sec.~\ref{XXZcalc2}).  Note that the initial state and the DE after the quench to the integrable part of $\hat H_F$ exhibit a peak at $k=\pi$, while such a peak is absent in the thermal predictions---there is no ``memory'' of the initial state distribution. The thermal predictions are almost $k$-independent because of the high energy density of the initial state in $\hat H_F$ (as in the experiments), which results in a high temperature $T_F$. The DE predictions can be seen to approach those of the GE with increasing $V_r$. As we argue next, the differences between these two ensembles in  Fig.~\ref{fig:XXZmomentum} are due to finite-size effects. They vanish in the thermodynamic limit.

\subsection{Thermalization}\label{XXZcalc2}

An important question that was not discussed in the main text in the context of the exact diagonalization calculations is whether the momentum distribution thermalizes, namely, whether the equilibrated momentum distribution function is that of a system in thermal equilibrium. Observables in integrable systems are expected to equilibrate but not thermalize, while in nonintegrable ones they are expected to thermalize \cite{DAlessio2015}. In order to determine whether the momentum distribution thermalizes, we first need to compute the GE prediction $m_k(\text{GE})=\text{Tr}[\hat m_{k}\hat \rho_\text{GE}]$ at the same energy and number of particles as in the time-evolved state. The density matrix of the grand canonical ensemble that describes thermalized observables is 
\begin{eqnarray}
\hat\rho_{\text{GE}}=\frac{\exp[-(\hat H_F-\mu_F \hat N)/T_F]}{\text{Tr}\{\exp[-(\hat H_F-\mu_F \hat N)/T_F]\}},
\end{eqnarray}
where $T_F$ and $\mu_F$ are found solving for the two equations: 
\begin{eqnarray}
\text{Tr}[\hat\rho_\text{GE}\hat H_F]&=&\text{Tr}[\hat \rho_{I}\hat H_F]\label{finalT},\\
\text{Tr}[\hat\rho_\text{GE}\hat N]  &=&\text{Tr}[\hat \rho_{I}\hat N].
\end{eqnarray}
Since our systems are always at half filling, $\mu_F=0$. 

We then compute a distance-to-thermalization metric
\begin{eqnarray}\label{eq:dge}
\delta_{\text{GE}}(\tau)=\sqrt{\displaystyle\frac{\sum_k \left[m_k(\tau)-m_k(\text{GE})\right]^2}{L/2}}.
\end{eqnarray} 

It is only in the thermodynamic limit that the diagonal ensemble (DE) predictions become identical to those of the GE in nonintegrable systems~\cite{rigol2014quantum,DAlessio2015}. Because of finite-size effects, $\delta_{\text{DE}}(\tau)$ and $\delta_{\text{GE}}(\tau)$ are different in our calculations away from integrability. To check that thermalization takes place in our nonintegrable systems, we also compute a distance between the diagonal and the grand canonical ensemble
\begin{eqnarray}\label{eq:ddege}
\delta(\text{DE-GE})=\sqrt{\displaystyle\frac{\sum_k \left[m_k(\text{DE})-m_k(\text{GE})\right]^2}{L/2}}
\end{eqnarray} 
and explore its behavior with changing system size.

\begin{figure}[t!]
    \includegraphics[width=1\columnwidth]{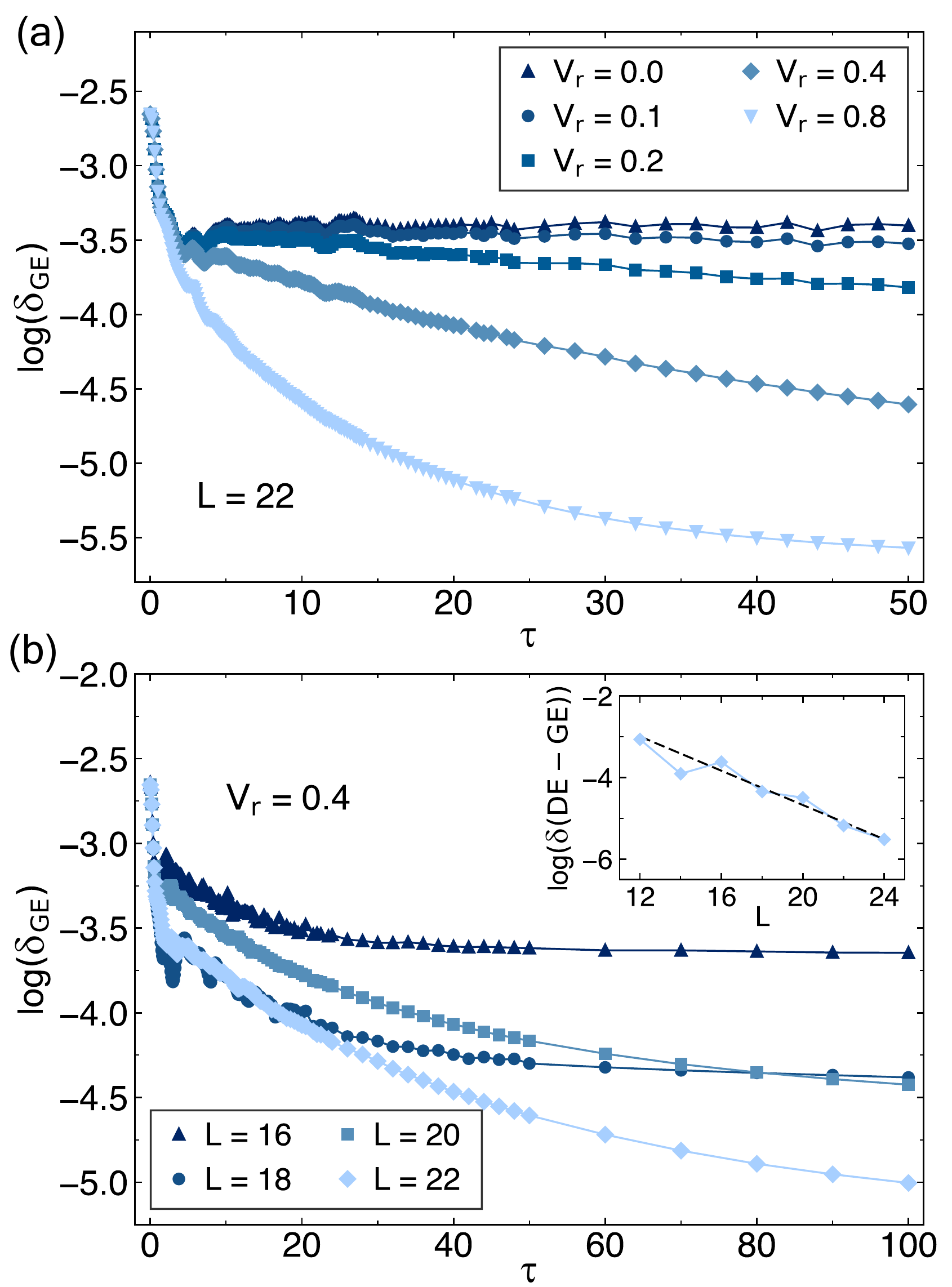}
    \caption{(a) RMS distance $\delta_{\text{GE}}(\tau)$ [see Eq.~\eqref{eq:dge}] versus $\tau$ for quenches in which $V=1.6$ and the integrability breaking $V_r$ takes different values. The results shown are for a system with $L=22$ sites. (B) RMS distance $\delta_{\text{GE}}(\tau)$ versus $\tau$ for quenches in which $V=1.6$, $V_r=0.4$, and for different lattice sizes ($L=$ 16, 18, 20, and 22). Inset in (b), distance between the diagonal and the grand canonical ensembles $\delta(\text{DE-GE})$ [see Eq.~\eqref{eq:ddege}] versus $L$ for $V=1.6$ and $V_r=0.4$ (as in the main panel). The black dashed line depicts exponential behavior.}
    \label{fig:XXZ_GE}
\end{figure}

In  Fig.~\ref{fig:XXZ_GE}(a), we plot the distance-to-thermalization $\delta_{\text{GE}}(\tau)$ for quenches with fixed $V=1.6$ but  different values of $V_r$. At any given time, one can see that $\delta_{\text{GE}}(\tau)$ is larger the closer the system is to integrability. One can also see that, for the largest value of $V_r$, $\delta_{\text{GE}}(\tau)$ converges to a nonvanishing value at long times. This is the result of finite-size effects. In Fig.~\ref{fig:XXZ_GE}(b), we plot $\delta_{\text{GE}}(\tau)$ for a fixed value of $V_r$ in chains with different number of sites. The plots show that the saturation value of $\delta_{\text{GE}}(\tau)$ at long times decreases with increasing system size. This suggests that, in the thermodynamic limit, the time-evolving momentum distribution function approaches the thermal prediction during the equilibration dynamics. Further evidence to support this expectation is presented in the inset in Fig.~\ref{fig:XXZ_GE}(b), in which we plot the distance between the diagonal and grand canonical ensemble predictions $\delta(\text{DE-GE})$ as a function of $L$. The results are consistent with $\delta(\text{DE-GE})$ vanishing exponentially with increasing $L$.

\subsection{Equilibration}\label{appequil}

\begin{figure}[t!]
  \includegraphics[width=1\columnwidth]{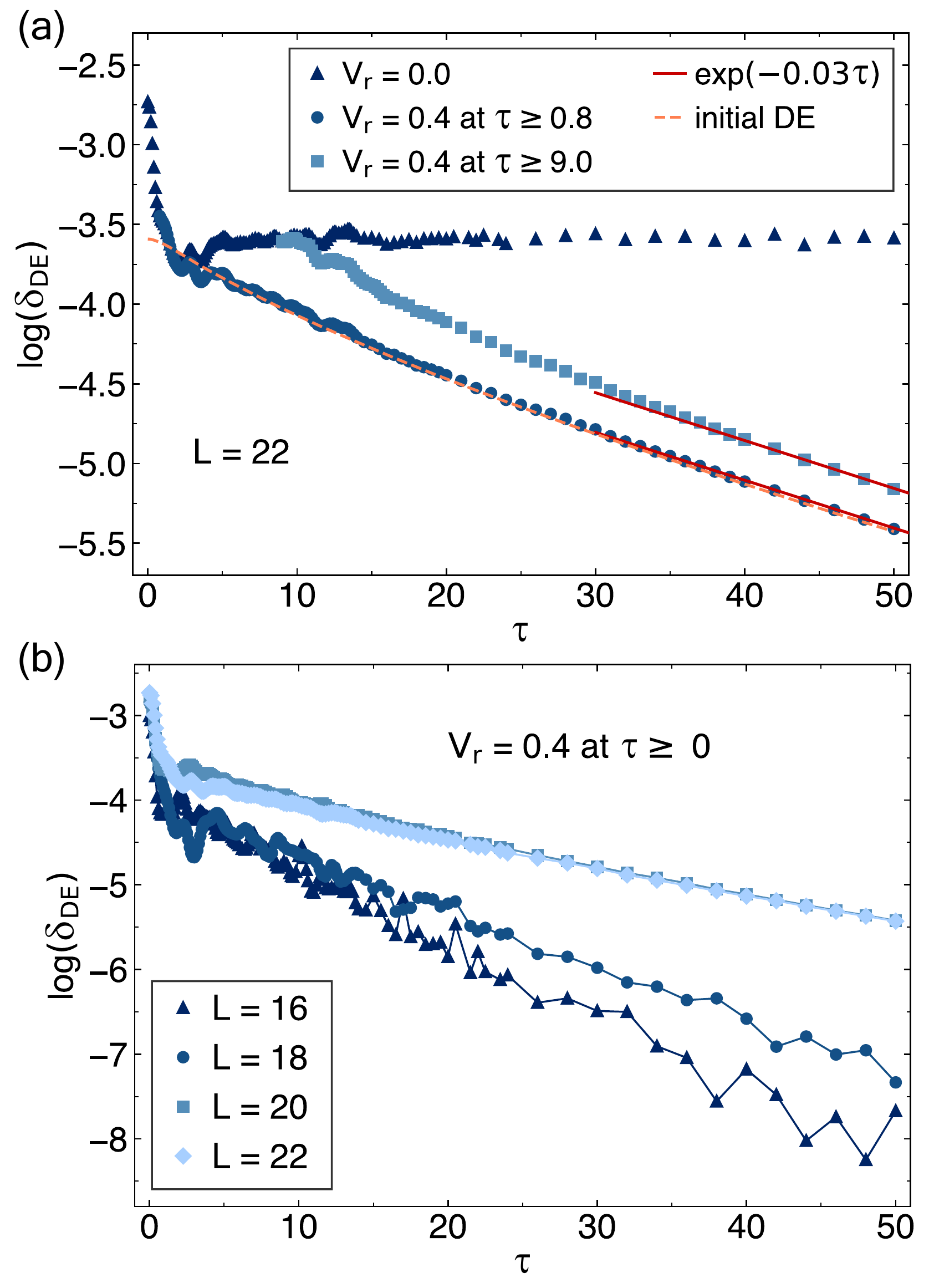}
    \caption{(a) RMS distance $\delta_{\text{DE}}(\tau)$ [defined in Eq.~\eqref{eq:dde}] versus $\tau$ for two-step quenches in which $V=1.6$, $V_r=0.4$, and $L=22$. The distance is computed, in all cases, from the DE result for the single quench. Results are shown when the second quench is carried out at $\tau=0.8$ and 9.0 following the first quench, and at $\tau=0$ when starting with the DE of the integrable system after the first quench (labeled as ``initial DE''). The solid lines at long times depict exponential behavior indicating the same relaxation rate in all cases. (b) Distance to equilibration in single quenches for different systems sizes $L$. These results show the finite-size effect on the relaxation rates.} 
    \label{fig:XXZtwoquench}
\end{figure}

Figure~\ref{fig:XXZtwoquench}(a) shows the distance to the diagonal ensemble of the single quench case, taking a state at two different times after the short-time dephasing following the first quench, and taking the diagonal ensemble after the first quench. They all can be seen to result in an exponential decay at long times to the diagonal ensemble of the single quench case. The exponentially decaying part exhibits nearly the same relaxation rate in the three curves (see fits). This shows that the thermalization rate is not affected by the choice of time to switch on $V_r$. The near-exponential relaxation can then be understood as generated by the time evolution of the DE of the integrable part of $\hat H_F$ under the nonintegrable $\hat H_F$.

The main limitation of our exact diagonalization results is, as mentioned before, finite-size effects. Figure~\ref{fig:XXZtwoquench}(b) shows the evolution of the distance-to-equilibration $\delta_{\text{DE}}(\tau)$ in the single quench protocol as one changes the system size ($L=16$, 18, 20, and $22$). The near-exponential relaxation is apparent in all cases, but the relaxation rate can be seen to be affected by finite-size effects.  Nevertheless, the trends manifest in the simulations qualitatively match those in the experiment, which has a far larger system size.

%\bibliographystyle{bibstyle.bst}
%\bibliography{newtonscradle.bib}


\end{document}